\documentclass[aps,pra,letterpaper,accepted=2020-09-17,longbibliography]{quantumarticle}
\pdfoutput=1

\usepackage{xkeyval}
\usepackage{etoolbox}
\usepackage{geometry}
\usepackage{xcolor}
\usepackage{fancyhdr}
\usepackage{ltxgrid}
\usepackage{ltxcmds}

\usepackage{graphicx}
\usepackage{epstopdf}
\usepackage{amsmath}
\usepackage{comment}
\usepackage{amssymb}

\usepackage{bm}
\usepackage[utf8]{inputenc}
\usepackage[english]{babel}
\usepackage[T1]{fontenc}
\usepackage{amsmath}
\usepackage{hyperref}
\usepackage{tikz}
\usepackage{lipsum}

\hypersetup{
    colorlinks=true,       % false: boxed links; true: colored links
    linkcolor=cyan,          % color of internal links
    citecolor=magenta,        % color of links to bibliography
    filecolor=magenta,      % color of file links
    urlcolor=cyan,           % color of external links
    runcolor=cyan
}
\newtheorem{theorem}{Theorem}
\newtheorem{observation}{Observation}

\newcommand{\beq}{\begin{equation}}
\newcommand{\eeq}{\end{equation}}
\newcommand{\beqnn}{\begin{equation*}}
\newcommand{\eeqnn}{\end{equation*}}
\newcommand{\bea}{\begin{eqnarray}}
\newcommand{\eea}{\end{eqnarray}}
\newcommand{\beann}{\begin{eqnarray*}}
\newcommand{\eeann}{\end{eqnarray*}}
\newcommand{\bes} {\begin{subequations}}
\newcommand{\ees} {\end{subequations}}

\newcommand{\ket}[1]{ | #1\rangle}
\newcommand{\bra}[1]{\langle #1 | }

\newcommand{\Tr}{\mathrm{Tr}}

\newcommand{\ignore}[1]{}

\newcommand{\poly}{\mathrm{poly}}

\newcommand{\ds}{\text{(de-signed)}}
\newcommand{\sh}{\text{(shifted)}}
\newcommand{\cV}{\mathcal{V}}
\newcommand{\cE}{\mathcal{E}}
\newcommand{\vol}{\textup{vol}}

\begin{document}
\title{De-Signing Hamiltonians for Quantum Adiabatic Optimization}

\author{Elizabeth Crosson}
\affiliation{Center for Quantum Information and Control (CQuIC),
Department of Physics and Astronomy , University of New Mexico, Albuquerque, NM 87131, USA}

\author{Tameem Albash}
\affiliation{Center for Quantum Information and Control (CQuIC),
Department of Physics and Astronomy , University of New Mexico, Albuquerque, NM 87131, USA}
\affiliation{Electrical and Computer Engineering, University of New Mexico, Albuquerque, NM 87131, USA}
\orcid{0000-0003-3916-3985}

\author{Itay Hen}
\affiliation{Information Sciences Institute, University of Southern California, Marina del Rey, California 90292, USA}
\affiliation{Department of Physics and Astronomy and Center for Quantum Information Science \& Technology, University of Southern California, Los Angeles, California 90089, USA}
\orcid{0000-0002-7009-7739}

\author{A. P. Young}
\affiliation{Department of Physics, University of California, Santa Cruz, California 95064, USA}

\begin{abstract}
\noindent Quantum fluctuations driven by non-stoquastic Hamiltonians have been conjectured to be an important and perhaps essential missing ingredient for achieving a quantum advantage with  adiabatic optimization. We introduce a transformation that maps every non-stoquastic adiabatic path ending in a classical Hamiltonian to a corresponding stoquastic adiabatic path by appropriately adjusting the phase of each matrix entry in the computational basis.  We compare the spectral gaps of these adiabatic paths and find both theoretically and numerically that the paths based on non-stoquastic Hamiltonians have generically smaller spectral gaps between the ground and first excited states, suggesting they are less useful than stoquastic Hamiltonians for quantum adiabatic optimization.   These results apply to any adiabatic algorithm which interpolates to a final Hamiltonian that is diagonal in the computational basis. 
\end{abstract}

\maketitle
%

%%%%%%%%%%%%%%%%%%%%%%%%%%%%%%%
\section{Introduction}

Quantum adiabatic optimization (QAO)~\cite{finnila_quantum_1994,Brooke1999,kadowaki_quantum_1998,farhi_quantum_2000,Santoro} is a  quantum algorithm that uses adiabatic evolution in Hamiltonian ground states to solve classical combinatorial optimization problems. The computational cost of the algorithm is given by the scaling with system size of the adiabatic condition~\cite{wannier:65,farhi:02,Elgart,Jansen:07,lidarGap}, which in turn depends on the minimum spectral gap encountered along the adiabatic path. 
While QAO is able to reproduce the known oracular speedup for quantum search~\cite{Roland:2002ul}, the method has not yet theoretically or empirically demonstrated an advantage over state-of-the-art classical algorithms in optimization~\cite{Young2008,Young:2010,hen:11,farhi:12}. 

The standard proposal for QAO uses an adiabatic path defined by a one parameter family of Hamiltonians $\{H(s)\}_{s\in [0,1]}$ that are all \emph{stoquastic}, meaning that $H(s)$ has real and non-positive off-diagonal matrix elements in the computational basis for each $s\in [0,1]$~\cite{bravyi2009complexity}.  Empirically it has been observed that this stoquastic QAO can in many cases be efficiently classically simulated by quantum Monte Carlo (QMC) methods~\cite{Isa2016}, meaning that QMC simulations are able to effectively follow the instantaneous ground state of the interpolating Hamiltonian with the same computational cost \emph{scaling} as QAO.
(QMC simulations can approximately sample equilibrium states in the computational basis and estimate the expectation values of local observables.)  There are now several rigorous results on polynomial-time QMC simulations for various classes of equilibrium states of stoquastic Hamiltonians~\cite{bravyi2009complexity,bravyi2015monte,crosson2016simulated,bravyi2017polynomial,crosson2018rapid,harrow2019classical}, however there are also counterexamples where QMC methods fail to converge efficiently~\cite{Hastings:2013kk,PhysRevA.94.042318,And2017}. 

In terms of computational complexity, estimating the ground state energy of a stoquastic Hamiltonian can be done in the complexity class AM which has a classical prover and classical verifier~\cite{bravyi2006complexity}, whereas estimating the ground energy of general local Hamiltonians is QMA-complete~\cite{Kitaev:book}.  Similarly, stoquastic adiabatic computation cannot be universal for quantum computing unless the Polynomial Hierarchy collapses~\cite{bravyi2006merlin}. Therefore, a significant open question is whether `non-stoquastic' Hamiltonians (which have positive or non-real off-diagonal elements in every choice of local basis~\cite{bravyi2009complexity,marvian:2018,Klassen2019twolocalqubit,2019arXiv190608800K}) could improve the performance of QAO~\cite{nonstoq2,PhysRevB.95.184416,PhysRevA.95.042321,nonstoq3,Alb2019,Dur2019,PhysRevA.98.062311}.  Adiabatic computation based on non-stoquastic Hamiltonians can be universal for quantum computing~\cite{aharonov2008adiabatic, Biamonte:07} and is not efficiently simulable by QMC due to the sign problem~\cite{Loh-PRB-90,Wiese-PRL-05,marvian:2018,gupta2019}.

In this work we define a locality-preserving mapping which takes every non-stoquastic QAO protocol to a corresponding stoquastic QAO protocol.  Considering various ensembles (dense matrices, signed graphs, and local Hamiltonians) we find that the non-stoquastic adiabatic paths have smaller spectral gaps than the corresponding stoquastic adiabatic paths, with high probability.   Using random matrix theory, spectral graph theory, and other analytical techniques we develop an explanation of this phenomenon based on the low-energy spectrum of stoquastic and non-stoquastic Hamiltonians. 

\section{Background and Overview}
In QAO, we assume that the global optima of discrete classical optimization problems are encoded in the ground states of a problem Hamiltonian $H_p$ that is diagonal in the computational basis~\cite{farhi_quantum_2000,Young2008,Young:2010,hen:11,hen:12,farhi:12}.
To reach a minimizing configuration of $H_p$, the system is initially prepared in the ground state of a non-diagonal `driver' Hamiltonian
$H_d$, with $\left[H_p, H_d \right] \neq 0$. 
The total Hamiltonian $H(s)$ of the system interpolates between $H_d$ and
$H_p$, with $s$ being the interpolation parameter.  (We suppress the dependence on $s$ unless otherwise noted.)  
If this process is carried out sufficiently slowly, the
adiabatic theorem of quantum mechanics~\cite{Kato:50,Jansen:07,lidarGap,Elgart}
ensures that the system will stay close to the ground state of the
instantaneous Hamiltonian throughout the evolution, such that one 
obtains a state close to the ground state of $H_p$ by the end of the interpolation.   
For the adiabatic approximation to hold it suffices for the running time 
of the algorithm to be inversely proportional to a low power of the minimum gap --- the difference between the second lowest and lowest eigenvalue of the Hamiltonian --- throughout the evolution~\cite{wannier:65,farhi:02,Elgart,Jansen:07,lidarGap}. 

In what follows, we focus solely on the gap of $H(s)$ and do not consider the role of different parameterizations of the interpolation parameter $s$ in terms of the physical time $t$. While appropriate reparameterizations of $s(t)$ can soften the dependence on the spectral gap, as in the adiabatic version of
the Grover search algorithm \cite{Roland:2002ul}, in general these techniques do not appear capable of removing the gap dependence associated with the adiabatic condition, unless the notion of adiabaticity is sacrificed altogether.  

To compare the performance of QAO driven by stoquastic and non-stoquastic fluctuations, 
we associate every non-stoquastic interpolating $n$-qubit Hamiltonian \hbox{$H = \sum_{i,j \in \{0,1\}^n} H_{ij} \ket{i}\bra{j}$} to a 
`stoquasticized' version of it, i.e., a stoquastic Hamiltonian whose sparsity structure (the location of non-zero off-diagonal matrix elements) is identical to that of the non-stoquastic one. We will consider two types of stoquastization. 
In the first, which we refer to as `de-signed' stoquastization, the de-signed stoquastic Hamiltonian is
\beq
H^\ds = \sum_{i \in \{0,1\}^n} H_{ii}|i\rangle \langle i| - \sum_{\substack{i,j \in \{0,1\}^n \\ i\neq j}} |H_{ij}| \ket{i}\bra{j} \,.
\eeq
The de-signing operation leaves the diagonal entries (corresponding to the classical cost function in QAO) unchanged, while adjusting the phases of the off-diagonal elements so that $H^\ds$ is stoquastic in the computational basis.  The transformation does not change the sparsity structure of the Hamiltonian, and in many cases the decomposition of $H^{\ds}$ into local Pauli terms can be determined from $H$ by inspection.  In the case where the off-diagonal entries arise purely from products of Pauli $X$ and $Y$ terms, we provide an algorithm for de-signing in Appendix~\ref{app:designing}.

A second form of stoquastization that we will consider applies to Hamiltonians with real matrix entries,
\beq \label{eqt:shift1}
H^\sh = \frac{1}{2} \left( H-|\max_{i\neq j} H_{ij}|  \sum_{i\neq j : H_{ij} \neq 0} |i\rangle \langle j|\right) \ ,  \,
\eeq
which we refer to as `shifting.'

We primarily consider shifting in the context of dense random matrices with i.i.d. entries, where it takes the form,
\beq
H^\sh = \frac1{2} \left( H-|\max_{i\neq j} H_{ij}| |s\rangle \langle s |\right)  \,. \label{eq:shiftdense}
\eeq
where $|s\rangle = \sum_{z \in \{0,1\}^n} |z\rangle$ is the unnormalized uniform superposition.  Similar to $H$, both $H^{\textrm{(de-signed)}}$ and $H^\sh$  define a path through the space of Hamiltonians that end at the same point $H_p$. We ask the natural question: which of the Hamiltonian paths has a larger minimum spectral gap, the non-stoquastic or the stoquasticized Hamiltonian?

To answer this question, we provide a series of results that cover different cases, all of which favor a larger spectral gap for the stoquasticized Hamiltonians.  (i) For a random $N\times N$ Hermitian matrix $H$ with independent and identically distributed (i.i.d.) entries drawn uniformly from $[-1,1]$ we prove that $\Delta_{H^\ds}$ is larger than $\Delta_H$ by a factor of $N^{3/2}$, with probability $1 - \exp(-\poly(N))$, where $\Delta$ denotes the energy gap between the ground state and first excited state.   (ii) For local Hamiltonians we first prove that if the total Hamiltonian $H$ is diagonal in the $X$ basis, then the stoquastization in the $Z$ basis, $H^\ds$, always obeys $\Delta_{H^\ds} \geq \Delta_H$.   (iii) To show that these properties extend throughout the interpolation path, we present numerical simulations of QAO up to $n = 22$ qubits which demonstrate that the fraction of QAO Hamiltonian paths with a larger minimum spectral gap than $H^\ds$ or $H^\sh$ rapidly goes to zero with $n$.  (iv) One may also consider quantum annealing algorithms which do not necessarily satisfy the adiabatic approximation, and for these we show by simulations of Schr{\"o}dinger time-evolution up to $n = 20$ qubits that the time-to-solution metric \cite{speedup} also favors stoquastic Hamiltonians.  (v) Finally, we apply techniques from spectral graph theory called signed graph Cheeger inequalities~\cite{atay2014cheeger} to derive new theorems relating the low-energy spectrum of signed (non-stoquastic) and unsigned (stoquasticized) versions of a graph.  Although these theorems do not fully capture the quantitative results that we find in the numerical simulations, we believe they present a valuable intuitive explanation for why spectral gaps of non-stoquastic Hamiltonians tend to be smaller than those of their stoquastic counterparts. We begin with a couple of preliminary observations.  

\subsection{Preliminary Observations}
Our first observation is that de-signing always reduces the ground state energy.

\begin{observation} For any $H$, $H^{\ds}$ with ground state energies $E_0$ and $E^{\ds}_0$ respectively, the ground state energies satisfy
\beq
E^\ds_0 \leq E_0. \label{eq:basicVariational}
\eeq
with equality holding if and only if there exists a unitary transformation $U$ such that $U H U^\dagger = H^\ds$, where $U$ is diagonal in the basis of the stoquastization.
\end{observation}
This bound [Eq.~\eqref{eq:basicVariational}] is proven by using the variational method with a ground state ansatz for $H^\ds$ that is formed from the absolute value of the components of the ground state of $H$.  (Details of the proof are given in Appendix~\ref{app:variationalGroundEnergy}.) In fact, this relationship between the ground state energies has previously been studied in a different but related context. In Ref.~\cite{PhysRevLett.111.100402,PhysRevB.97.125153} the authors consider many-body systems consisting of particles hopping on a graph, and interacting by a potential (i.e., Hubbard models on a general graph).  In our terminology, they prove that if the hopping part of the Hamiltonian is stoquastic then bosons always have a \emph{strictly} lower energy than fermions.  However, if the hopping Hamiltonian is non-stoquastic then fermions can have a lower ground energy (note that the spectral gaps of these systems are not considered in~\cite{PhysRevLett.111.100402,PhysRevB.97.125153}). 

While the de-signed Hamiltonian $H^\ds$ always has a lower ground state energy, it is easy to generate examples in which the first excited energies satisfy $E^\ds_1 < E_1$.  Therefore, showing that $\Delta^\ds \geq \Delta$ with high probability requires a more sophisticated understanding of how much the first excited energy can be decreased by de-signing.

Note that we can always shift our non-stoquastic Hamiltonian to obtain a positive semidefinite matrix $G= \|H\| I  - H$, with the ground state of $H$ becoming the principal eigenvector of $G$ (here $\|H\|$ denotes the operator norm of $H$). 
Therefore we can reformulate our problem in terms of the positive semidefinite matrix $G$ and its counterpart $G^+$ that is the non-negative matrix formed by taking the entrywise absolute values of $G$.  We now compare the spectral gap between the largest and next largest eigenvalues of $G$ and $G^+$, which leads to the following observation:
\begin{observation}
Let $G, G^+$ be as above.  Then $\|G^+\| \geq \|G\|$ by Eq.~\eqref{eq:basicVariational}, but $\Tr(G) = \Tr(G^+)$, so at least some of the eigenvalue gaps in the spectrum of $G$ must be smaller than the corresponding gaps in $G^+$.  
\end{observation}
In the remainder of this work we will provide theoretical and numerical evidence that this ``compression'' of the spectrum is concentrated in the low-energy sector.

Finally, our third preliminary observation pertains to  the first-order perturbative correction that arises when a Hamiltonian $H_0$ that is stoquastic (in, say, the computational basis) is perturbed by a Hamiltonian $V$ with non-negative entries in the computational basis (therefore $V$ perturbs $H$ towards becoming non-stoquastic).   If $|\psi_0^0\rangle$ is the ground state of $H_0$ with energy $E_0^0$, then $|\psi_0^0\rangle$ has all non-negative amplitudes in the computational basis.  The perturbed Hamiltonian $H_s = H_0 + s V$ has ground energy $E_0^0 + s \langle \psi_0^0 | V | \psi_0^0\rangle$ to first order in $s$.    Similarly, if $|\psi_0^1\rangle$ is the first excited state of $H_0$ with energy $E_0^1$, then to first order in $s$ the first excited energy of $H_s$ is $E_s^1 = E_0^1 + s \langle \psi_0^1 | V | \psi_0^1\rangle$.  Therefore, up to first order in $s$ we have
$$
\Delta_s = \Delta_0 + s \left (\langle \psi_0^1 |V | \psi_0^1\rangle - \langle \psi_0^0 | V |\psi_0^0\rangle \right)
$$
The fact that $V$ has non-negative entries, and $|\psi_0^0\rangle$ has non-negative amplitudes, means that the $\langle \psi_0^0 | V | \psi_0^0\rangle$ is non-negative (it is a sum of non-negative numbers).  Meanwhile, $\langle \psi_0^1 | V | \psi_0^1\rangle$ is a sum of numbers with opposite signs.   This suggests that a perturbation in the direction of becoming non-stoquastic increases the ground energy, and may either increase or decrease the first excited energy, but likely by a lesser amount (due to the cancellation) unless the unperturbed first excited state aligns with the perturbation in a finely tuned way. Such an effect was demonstarted in Ref.~\cite{Alb2019}, where an example was provided where a non-stoquastic intermediate Hamiltonian appears to mitigate (the study was limited to $n \leq 24$) a perturbative crossing \cite{Ami2009} that the de-signed Hamiltonian could not.  

\section{Rigorous results}

We rigorously compare general Hamiltonians to their de-signed and shifted counterparts in three complimentary regimes: random dense matrices with i.i.d. entries (compared to their shifted counterparts), Hamiltonians which are diagonal in the $X$ basis compared to their counterparts which have been de-signed in the $Z$ basis, and non-stoquastic Hamiltonians corresponding to signed graph Laplacians, which are compared to their de-signed counterparts (i.e., the underlying unsigned graphs).  
\subsection{Random matrices} \label{sec:RandomMatrices}

In this section we apply results from random matrix theory to compare the spectral gap (between the largest two eigenvalues) of $N \times N$ random real symmetric matrices to the corresponding spectral gap of their shifted counterparts (which form an ensemble of symmetric matrices with non-negative entries).  The random matrices we consider have i.i.d. entries, and therefore these matrices are dense with non-zero entries, almost surely.  For the real symmetric matrices we will consider a distribution $\mathcal{W}$ of non-Gaussian Wigner matrices (which are defined in general to have entries distributed with mean 0 and finite variance).  In our case the matrices in $\mathcal{W}$ are defined to have entries that are uniformly distributed in $[-1,1]$, in order to use $\max_{i,j} |H_{i,j}| = 1$ in Eq.~\eqref{eq:shiftdense}.  

We will compare these matrices to a distribution $\mathcal{W}^+$ of random matrices with nonnegative entries drawn uniformly from $[0,1]$ --- these would be the `shifted' matrices.  The key observation that enables the analysis is that these two matrix distributions are related by a rank 1 shift matrix, $A_{ij} = 1$ for all $i,j = 1,...,N$.  Specifically, for every random matrix $\hat{W}^+$ drawn from $\mathcal{W}^+$ there exists a $\hat{W}$ drawn from $\mathcal{W}$ such that
\begin{equation}
\hat{W}^+ = A + \frac{1}{2} \hat{W} \quad , \quad \textrm{Pr}_{\mathcal{W}^+}(\hat{W}^+) = \textrm{Pr}_{\mathcal{W}}(\hat{W}) ,\label{eq:finiterank}
\end{equation}
where $\textrm{Pr}_{\mathcal{W}^+}$ denotes the probability of drawing the matrix $W^+$ from the ensemble $\mathcal{W}^+$, and $\textrm{Pr}_{\mathcal{W}}(\hat{W})$ is defined similarly.  We note that similar arguments involving finite rank perturbations have previously appeared in the random matrix literature~\cite{feral2007largest}, however our application to comparing spectral gaps of stoquastic and non-stoquastic Hamiltonians is novel. 

We will apply a large deviation bound that can be found as corollary 2.3.5 in Ref.~\cite{tao2012topics}, which we reproduce here for completeness.\\
\emph{Lemma.} Suppose
that the entries $\hat{w}_{ij}$ of $W$ are independent, have mean zero, and are uniformly bounded in magnitude by 1. Then there exist absolute constants $C,c > 0$ such that
\begin{equation}
\textrm{Pr}_{\hat{W}} [ \|\hat{W} \| > \ell \sqrt{N} ] \leq C e^{-c \ell N} \label{eq:tailbound}
\end{equation}
for all $\ell \geq C$.  This result states that the largest eigenvalue of $\hat{W}$ is $\mathcal{O}(\sqrt{N})$ with probability exponentially close to 1.  Even tighter bounds than this are known, which show that $\|\hat{W}\|$ lies in the interval \hbox{$[2 \sqrt{N} - \mathcal{O}(N^{1/6} , 2 \sqrt{N} + \mathcal{O}(N^{1/6})]$}, but Eq.~\eqref{eq:tailbound} has the advantage of a simple tail bound and suffices for our purposes. 

Using Eqs.~(\ref{eq:finiterank}-\ref{eq:tailbound}) together with Weyl's inequality~\cite{horn1990matrix,weyl} allows us to show that the spectral gap of $W^+$ is $\Omega(N)$\footnote{The notation $\Omega(f(n))$ means ``asymptotically no smaller than $f(n)$''.  } with high probability.  Let the eigenvalues of $W^+$ be \hbox{$\lambda^+_1 \geq ... \geq \lambda^+_N$}, the eigenvalues of $A$ be $a_1 = N \geq a_2 = 0\geq .... \geq a_N = 0$, and the eigenvalues of $W$ be \hbox{$\lambda_1 \geq ... \geq \lambda_N$}.  Then Weyl's inequality implies
$$
\lambda^+_1 \geq a_1  +\frac{1}{2}\lambda_N  \geq N + \mathcal{O}(\sqrt{N})
$$
and also 
$$
\lambda^+_2 \leq a_2 + \frac{1}{2}\lambda_1 \leq 0 + \mathcal{O}(\sqrt{N})
$$
therefore we have $\Delta_{W^+} = \lambda^+_1 - \lambda^+_2 \geq N + \mathcal{O}(\sqrt{N})$.  Meanwhile, from the famous Wigner semicircle distribution~\cite{wigner1958distribution, tao2012topics} we have that $\Delta_{W} = \lambda_1 - \lambda_2 =  \mathcal{O}(N^{-1/2})$.  Therefore, in this normalization as $N\to  \infty $ we have $\Delta_W \to 0$ and $\Delta_{W^+} \to \infty$. 
\subsection{Gaps of $X$-stoquastic matrices}

We next show that if $H$ that is diagonal in the Pauli-$X$ basis (and is therefore obviously stoquastic) then the spectral gap is not reduced by de-signing in the $Z$ basis. To show this, let us denote the two lowest eigenvalues of $H$ by $\lambda_0$ and $\lambda_1$ and the two  lowest eigenvalues of $H^{\ds}$ by  $\lambda_0^\ds$ and $\lambda_1^\ds$.
In our proof, we shall use the following representation of $H$~\footnote{Note that the number of free parameters $\alpha,\alpha_i,\alpha_{ij},\ldots$ is precisely $2^n$, where $n$ is the number of spins, hence the representation is unique.} :
\beq\label{eq:Hx}
H=\alpha + \sum_i \alpha_i X_i +  \sum_{i<j} \alpha_{ij} X_i X_j +  \sum_{i<j<k} \alpha_{ijk} X_i X_j X_k + \ldots
\eeq
We further decouple the magnitude of the various $\alpha$ parameters from their signs and denote
\beq
\alpha_i=|\alpha_i| (-1)^{a_i}, \alpha_{ij}=|\alpha_{ij}| (-1)^{a_{ij}}, \ldots
\eeq
with $a_i,a_{ij},a_{ijk},\ldots \in \{ 0,1\}$. The de-signed version of $H$ is
 \bea
 H^\ds&=&\alpha - \sum_i |\alpha_i| X_i -  \sum_{i<j} |\alpha_{ij}| X_i X_j \nonumber\\
 &-& \sum_{i<j<k} |\alpha_{ijk}| X_i X_j X_k - \ldots
 \eea
 and since both $H$ and $H^\ds$ are diagonal in the $x$-basis, we can replace the Pauli operators $X_i$ with their bit representations $X_i=(-1)^{x_i}$, where $x_i\in \{ 0,1\}$.
 
 For convenience and without loss of generality we will fix the constant $\alpha$ to be
 \beq
 \alpha = \sum_i |\alpha_i| + \sum_{i<j} |\alpha_{ij}| + \sum_{i<j<k} |\alpha_{ijk}|+ \ldots
 \eeq
 so that $H$ and $H^\ds$ can be written as:
 \bea
 H&=&\sum_i |\alpha_i| \left(1-(-1)^{a_i\oplus x_i}\right]\\
 &+&  \sum_{i<j} |\alpha_{ij}| \left(1-(-1)^{a_{ij}\oplus x_i\oplus x_j}\right)\nonumber\\
 &+&  \sum_{i<j<k} |\alpha_{ijk}| \left(1-(-1)^{a_{ijk} \oplus x_i\oplus x_j \oplus x_k}\right) + \ldots)\nonumber
 \eea
 and similarly 
 \bea
  H^\ds&=&\sum_i |\alpha_i| \left(1-(-1)^{x_i}\right) \\
  &+&  \sum_{i<j} |\alpha_{ij}| \left(1-(-1)^{x_i\oplus x_j}\right) \nonumber\\
  &+& \sum_{i<j<k} |\alpha_{ijk}| \left(1-(-1)^{x_i\oplus x_j \oplus x_k} \right) + \ldots\nonumber
 \eea
 where $\oplus$ denotes bitwise XOR.
 It is easy to see that $H^\ds$ is minimized by setting $x_i=x^{G}_i=0$ giving $\lambda_0^\ds=0$. 
 Further denoting the ground and first excited state configurations of $H$ by $\tilde{x}^{G}_i$ and  $\tilde{x}^{E}_i$, respectively, and the first excited state configuration of $H^\ds$ by  $x^{E}_i$, let us now prove that \hbox{$\Delta^\ds=\lambda^\ds_1 \geq \Delta=\lambda_1-\lambda_0$}.
 The eigenvalue $\lambda_1^\ds$ evaluates to
 \bea\label{eq:lambda1}
\lambda_1^\ds&=&\sum_i |\alpha_i| \left(1-(-1)^{x^{E}_i}\right) \\
&+&  \sum_{i<j} |\alpha_{ij}| \left(1-(-1)^{x^{E}_i\oplus x^{E}_j}\right) \nonumber\\
  &+& \sum_{i<j<k} |\alpha_{ijk}| \left(1-(-1)^{x^{E}_i\oplus x^{E}_j \oplus x^{E}_k} \right) + \ldots\nonumber
 \eea
 and similarly 
  \bea
\Delta&=&\sum_i |\alpha_i| (-1)^{a_i}\left( (-1)^{\tilde{x}^{G}_i}-(-1)^{\tilde{x}^{E}_i} \right) \nonumber \\
&+&  \sum_{i<j} |\alpha_{ij}| (-1)^{a_{ij}}\left( (-1)^{\tilde{x}^{G}_i\oplus\tilde{x}^{G}_j}-(-1)^{\tilde{x}^{E}_i\oplus\tilde{x}^{E}_j} \right) \nonumber\\
  &+& \sum_{i<j<k} |\alpha_{ijk}| (-1)^{a_{ijk}}\left((-1)^{\tilde{x}^{G}_i\oplus\tilde{x}^{G}_j\oplus\tilde{x}^{G}_k} \right. \nonumber \\
  &-& \left.(-1)^{\tilde{x}^{E}_i\oplus\tilde{x}^{E}_j\oplus\tilde{x}^{E}_k}\right)  \ldots 
 \eea
Rearranging $\Delta$ above, we get:
  \bea
\Delta&=&\sum_i |\alpha_i| (-1)^{\tilde{a}_i}\left( 1-(-1)^{\delta_i} \right) \\\nonumber
&+&  \sum_{i<j} |\alpha_{ij}| (-1)^{\tilde{a}_{ij}}\left(1-(-1)^{\delta_i \oplus \delta_j} \right) \\\nonumber
  &+& \sum_{i<j<k} |\alpha_{ijk}| (-1)^{\tilde{a}_{ijk}}\left(1-(-1)^{\delta_i \oplus \delta_j\oplus \delta_k}\right) + \ldots
 \eea
 where $\delta_i=\tilde{x}^{G}_i\oplus \tilde{x}^{E}_i$, $\tilde{a}_i=a_i \oplus \tilde{x}^{G}_i, \tilde{a}_{ij}=a_{ij} \oplus \tilde{x}^{G}_i \oplus \tilde{x}^{G}_j, \tilde{a}_{ijk}=a_{ijk}+\tilde{x}^{G}_i\oplus\tilde{x}^{G}_j\oplus\tilde{x}^{G}_k$ and so forth.
 
 We can upper-bound  the gap $\Delta$ by replacing the first excited state configuration by a (possibly) different configuration (which may have a higher cost) given by $\tilde{x}^{G}_i \oplus x^{E}_i$.
 This gives $\delta_i=x^{E}_i$ and a bound on the gap, $\Delta \leq \tilde{\Delta}$, where 
 \bea
 \tilde{\Delta} &=& \sum_i |\alpha_i| (-1)^{\tilde{a}_i}\left( 1-(-1)^{x^{E}_i} \right) \\\nonumber
 &+&  \sum_{i<j} |\alpha_{ij}| (-1)^{\tilde{a}_{ij}}\left(1-(-1)^{x^{E}_i\oplus x^{E}_j} \right) \nonumber\\
  &+& \sum_{i<j<k} |\alpha_{ijk}| (-1)^{\tilde{a}_{ijk}}\left(1-(-1)^{x^{E}_i\oplus x^{E}_j\oplus x^{E}_k}\right) + \ldots \nonumber
 \eea
 Comparing $\tilde{\Delta}$ above with the expression obtained for $\lambda_1^\ds$, Eq.~(\ref{eq:lambda1}), we conclude that $\tilde{\Delta} \leq \lambda_1^\ds$ and so
 \beq
 \Delta^\ds =\lambda_1^\ds \geq \tilde{\Delta} \geq \Delta \,.
 \eeq
 Thus, the gaps of $X$-diagonal matrices are never decreased by de-signing in the $Z$ basis.

\subsection{Spectral graph theory}

In this section, we consider signed graph Laplacians, which serve as an interesting class of non-stoquastic Hamiltonians  (including some sparse and local Hamiltonians), and enable the use of signed graph Cheeger inequalities~\cite{atay2014cheeger} to derive new theorems that relate the low-energy spectrum of a signed (non-stoquastic) graph Laplacian to that of its unsigned (stoquastic) counterpart.  This work belongs to a line of progress in adapting discrete Cheeger inequalities~\cite{chung1997spectral} and related bounds to the setting of quantum Hamiltonians~\cite{al2010energy, jarret2014adiabatic,crosson2017quantum}, including the recent adaptation of signed graph Cheeger inequalities~\cite{gournay2016isoperimetric, martin2017frustration, atay2014cheeger} to non-stoquastic Hamiltonians~\cite{jarret2018hamiltonian, jarret2018quantum}.     The bounds we obtain are not probabilistic, but the drawback is that they also depend on some geometric properties of the associated ground states.  Roughly speaking, the cases where the non-stoquastic spectral gap can be larger than the stoquastic spectral gap correspond to highly localized non-stoquastic ground states that are only nonzero in a small sector of the local basis.  Despite this limitation, we present these results in part because they lead to an appealing intuitive explanation for the general observations we make about stoquastic \emph{vs.} non-stoquastic spectra.

The Laplacian of an unsigned graph $G = (\cV , \cE)$, where $\cV = \{v_1,...,v_N\}$ is a set of vertices and $\cE \subseteq \cV \times \cV$ is a set of edges between those vertices, is an $N\times N$ symmetric matrix $L$ with the degree of each vertex along the diagonal, and off-diagonal matrix entries that are zero except for $L_{ij} = -1$ corresponding to edges $(v_i,v_j) \in \cE$ (see appendix~\ref{app:sgt} for full details including the generalization to graphs with weighted edges).   Therefore these unsigned graph Laplacians always correspond to stoquastic Hamiltonians, with vertices corresponding to basis states and weighted edges correspoinding to off-diagonal matrix elements.  In a signed graph each edge $(v_i, v_j)$ is associated with a signature $\sigma_{ij} \in \{+1,-1\}$, and the signed Laplacian matrix $L^\sigma$ has off-diagonal entries $L^\sigma_{ij} = -\sigma_{ij}$, while the diagonal entries are again given by the (unsigned) degrees of the vertices.  Since $L^\sigma$ can now contain positive off-diagonal elements, such signed graph Laplacians can be non-stoquastic Hamiltonians.  Our goal is to compare the eigenvalues $\{\lambda^\sigma_i\}_{i = 1}^{|V|}$ of a graph with signature $\sigma$, enumerated in nondecreasing order, to the eigenvalues $\{\lambda^+_i\}_{i = 1}^{|V|}$ of its unsigned counterpart.  

Since signed graph Laplacians can contain a sign problem, one may wonder about the conditions under which this problem might be cured.  The natural class of transformations to consider in this context are of the form $S^\dagger L^\sigma S$, where $S$ is a signature matrix (a diagonal unitary with real entries). For this class of transformations it is possible to completely characterize the cases in which the non-stoquasticity can be cured, and these signed graphs are called ``balanced.''  A signed graph is balanced if and only if the product of $\sigma_{ij}$ around every cycle in the graph is positive. 
Another equivalent characterization is that every subset $S \subseteq V$ of the graph can be partitioned into two sets $S = S_1 \cup S_2$, with no negative edges ($\sigma_{ij} = -1$) within the sets, and no positive edges between them.  If we denote the number of positive edges between $S_1$ and $S_2$ by $|\cE^+(S_1 , S_2)|$, and the number of negative edges within the sets by $ |\cE^- (S_1)|$ and $ |\cE^- (S_2)|$ respectively, then a balanced graph satisfies $|\cE^+(S_1 , S_2)|+ |\cE^- (S_1)|  + |\cE^-(S_2)| =0$.  

It turns out that this idea of bipartitioning a subset of vertices $S\subseteq{V}$ into subsets $S_1$, $S_2$ leads to a useful quantity $F(S)$ called the (intensive) frustration index (note that we modify the notation of Ref.~\cite{atay2014cheeger} to better suit a comparison of signed and unsigned graphs), 
\begin{align}
F(S) := \frac{1}{\vol(S)} \cdot \min_{\substack{S_1 \cap S_2 = \emptyset \\S_1 \cup S_2 = S}} \bigl (2 |\cE^+(S_1,S_2)|\nonumber\\ + |\cE^-(S_1)| + |\cE^-(S_2)| \bigr ),\label{eq:frustrationIndex}
\end{align}
where $\vol(S)$ is the sum of the vertex degrees in $S$. 
Through the signed Cheeger inequalities we will see that the frustration index has an essential role in characterizing the low-energy spectrum of the graph.

For unsigned graphs, Cheeger's inequality and its ``multi-way'' generalizations characterize the low energy spectrum in terms of a quantity called the \emph{expansion}, which detects bottlenecks in the graph.  For any nonempty subset $S \subseteq V$ the \emph{expansion} is defined by
$$
\Phi(S) = \frac{|\cE(S,\bar{S})|}{\vol(S)}.
$$
where $|\cE(S,\bar{S})|$ denotes the number of (unsigned) edges leaving $S$.  The expansion $\Phi(S)$ is also sometimes called the ``bottleneck ratio'', since it measures the size of the boundary of the set to the size of its interior.   For any $k \geq 1$ the $k$-th order Cheeger constant is 
\begin{equation}
h^+_k= \min_{\substack{S_1,...,S_k \subseteq  \cV \\S_i \cap S_j = \emptyset}}  \max \left \{\Phi(S_1), ... , \Phi(S_k) \right\}. \label{eq:multiStoqCheegerDef1}
\end{equation}
In words, $S_1,...,S_k$ is the sub-partition that minimizes the expansion of all the subsets, and $h^+_k$ is the largest expansion amongst those subsets in the sub-partition.   Cheeger's inequality relates $h_k$ to the $k$-eigenvalue,
\begin{equation}
\frac{\lambda^+_k}{2} \leq h_k^+ \leq C k^3 \sqrt{2 D_{\max} \lambda_k^+},\label{eq:multiStoqCheeger1}
\end{equation}
where $C$ is a constant and $D_{\max}$ is the maximum (unsigned) degree of any vertex.  The conclusion from Eq.~\eqref{eq:multiStoqCheeger1} is that the low-energy spectrum of an unsigned graph (and in fact any stoquastic Hamiltonian) is characterized by sub-partitioning the ground space into subsets of minimal expansion: one has $k$ small eigenvalues if and only if there are at least $k$ disjoint subsets with bottlenecks.  

To generalize Cheeger's inequality to signed graphs, one defines the following signed Cheeger constant 
\beq
h^\sigma_k= \min_{\substack{S_1,...,S_k \subseteq  \cV \\S_i \cap S_j = \emptyset}}  \max \left \{F(S_1)  +\Phi(S_1) , ... , F(S_k) + \Phi(S_k) \right\} \ , \label{eq:higherCheegerDef}
\eeq
and then the generalization of Eq.~\eqref{eq:multiStoqCheeger1} is
\begin{equation}
\frac{\lambda^\sigma_k}{2} \leq h_k^\sigma \leq C k^3 \sqrt{2 D_{\max} \lambda_k^\sigma}.\label{eq:higherSignedCheeger1}
\end{equation}
Therefore a signed graph has at least $k$ small eigenvalues above the ground state if and only if it can be sub-partitioned into at least $k$ disjoint subsets that all have low expansion and a small frustration index.  

\begin{figure}[h]
\begin{center}
\includegraphics[width=0.47 \textwidth]{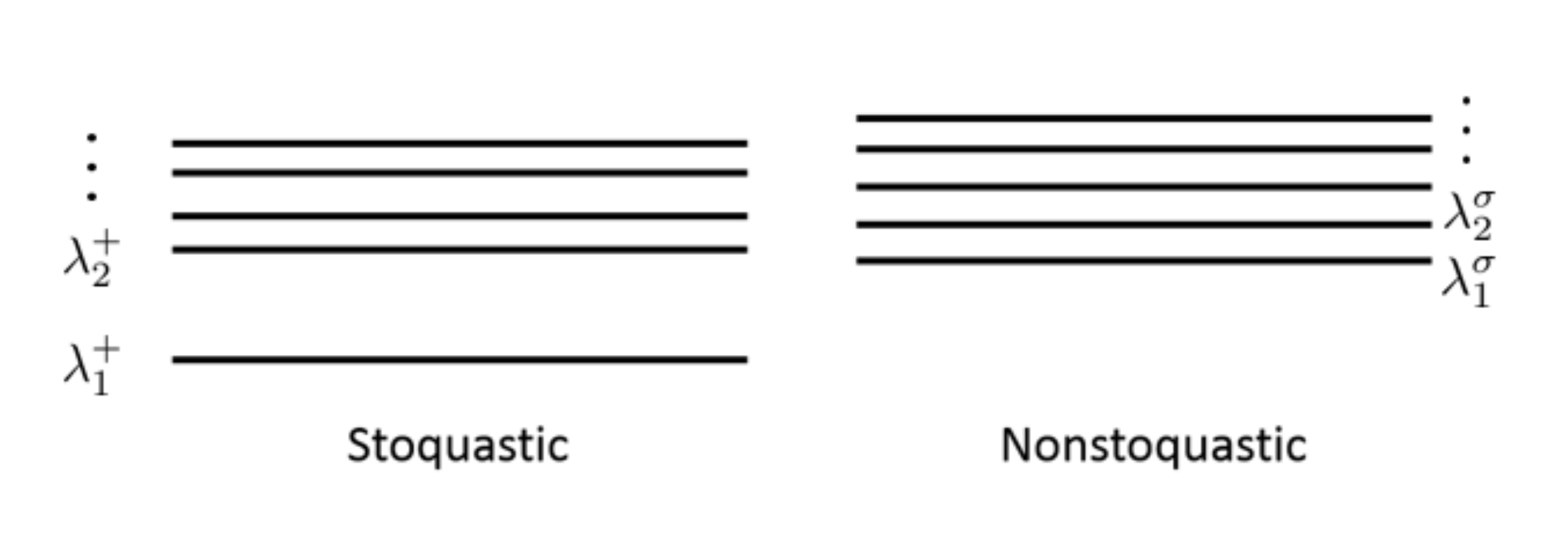}
\caption{An intuitive general picture of the energy spectrum of a stoquastic Hamiltonian compared to its de-signed counterpart. } \label{fig:intuition}
\end{center}
\end{figure}

The Cheeger inequalities already provide an argument why we might generically expect non-stoquastic Hamiltonians to have smaller spectral gaps. Consider a stoquastic spectrum and the associated Cheeger subsets, $\lambda^+_2$ with $S_1^{(2)},S_2^{(2)}$, $\lambda^+_3$ with $S_1^{(3)},S_2^{(3)},S_3^{(3)}$, etc.  These subsets all try to minimize their expansion.  What upper bounds on the non-stoquastic eigenvalues do we get by considering these subsets?  The upper bound we obtain on $\lambda^\sigma_1,\lambda^\sigma_2,\lambda^\sigma_3$ is now increased by the frustration index of these subsets.  But note that $S_1^{(2)},S_2^{(2)}$ will usually be bigger subsets than $S_1^{(3)},S_2^{(3)},S_3^{(3)}$, and so in the typical case $S_1^{(2)},S_2^{(2)}$ will have a larger frustration index.  This rough arguments suggests that the ground state energy is pushed up most by phase frustration, the first excited energy is pushed up slightly less, etc.  We illustrate this intuition in Fig.~\ref{fig:intuition}.  This then provides an explanation for why non-stoquastic Hamiltonians tend to have smaller spectral gaps.

\begin{figure}
\begin{center}
\includegraphics[width=.4\textwidth]{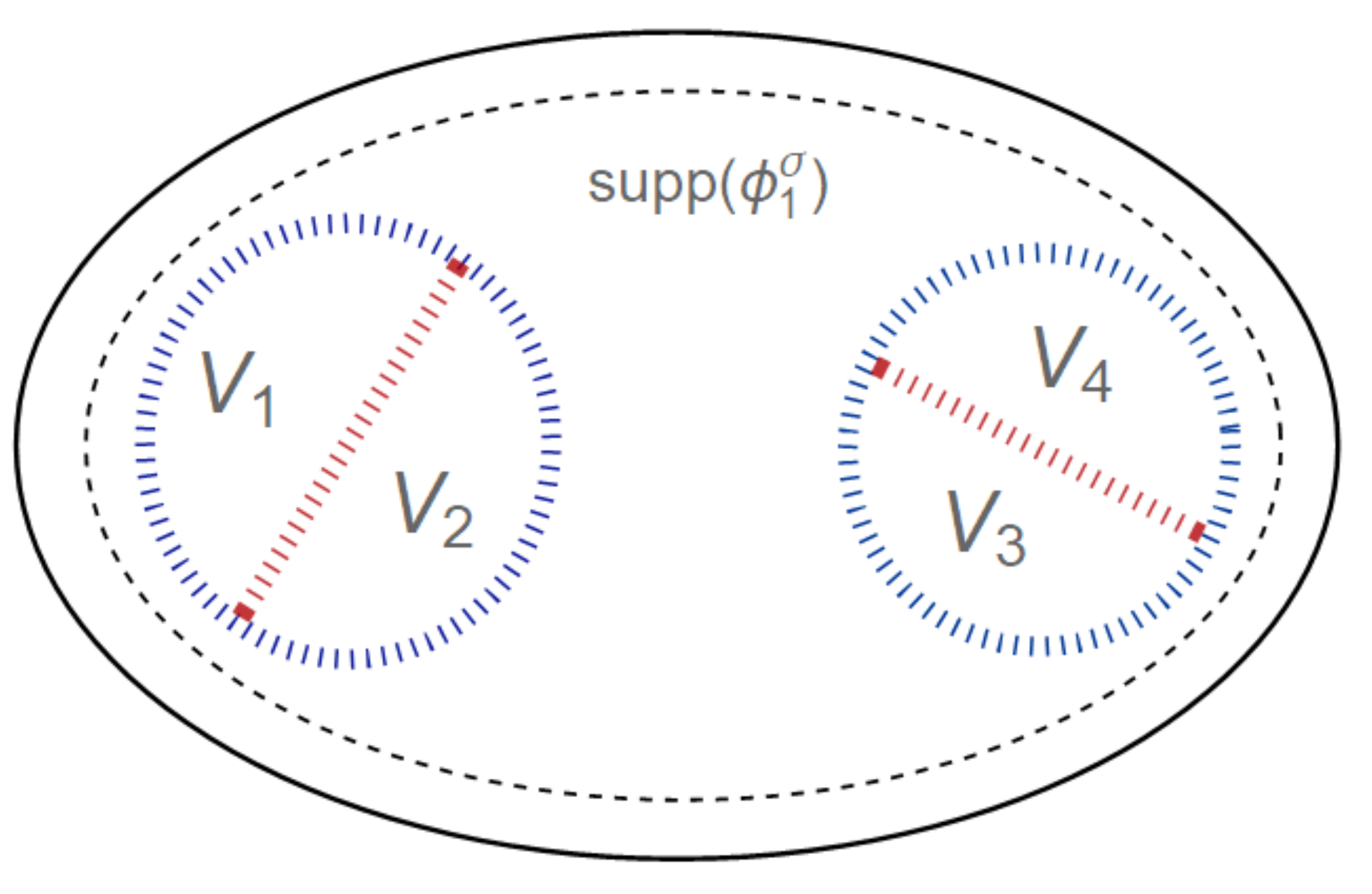}
\caption{\label{fig:UnsignedHigherCheeger}An illustration of our proof strategy for Theorem~\ref{thm:main}.  Details are provided in Appendix~\ref{app:sgt}. The figure shows the support of the non-stoquastic ground state $\phi_1^{\sigma}$ and two subsets $S'_1 = V_1 \cup V_2$ and $S'_2 = V_3 \cup V_4$ with a small bottleneck ratio and a small frustration index that is determined by the phase frustration that already exists in the non-stoquastic ground state. Together with the signed and unsigned Cheeger inequalities, we show these subsets provide a decent upper bound on the non-stoquastic first excited energy.}
\end{center}
\end{figure}

Next we apply these Cheeger inequalities to prove a new result that relates the spectral gap of a signed graph with its de-signed counterpart.  We have already seen in Eq.~\eqref{eq:basicVariational} that $\lambda_1^+ \leq \lambda_1^\sigma$ for all $\sigma$ (an unbalanced signature $\sigma$ cannot decrease the ground energy), and so we now seek to obtain an upper bound on $\lambda_2^\sigma$.
 To do this we can construct a variational excited state for $L^\sigma$ as follows.  From $\lambda_1^\sigma$ and Eq.~\eqref{eq:higherSignedCheeger1} we can upper-bound the expansion and the frustration index of the subset $\Omega$ corresponding to the support of the non-stoquastic ground state. 
  From $\lambda^+_2$ and the unsigned Cheeger inequality Eq.~\eqref{eq:multiStoqCheeger1} we also know there is a subset $S_1$ and $S_2$ with minimal expansion. Therefore we can upper bound  the expansion of the intersections $S_1' = S_1 \cap \Omega$ and $S_2' = S_2 \cap \Omega$ using Eq.~\eqref{eq:multiStoqCheeger1}, and upper bound the frustration with respect to a bipartition $S_1' = V_1 \cup V_2$, $S_2' = V_3 \cup V_4$ inherited from the non-stoquastic ground state (see fig.~\ref{fig:UnsignedHigherCheeger}).  These subsets are used to obtain the upper bound on $\lambda_2^\sigma$ in the following theorem.  
\begin{theorem}\label{thm:main}
\textit{ Let $L^\sigma$ be a signed graph Laplacian with ground state $\phi_1^\sigma$ with energy $\lambda^\sigma_1$.  Define $\Omega := \{v \in V : \phi_1^\sigma(v) \neq 0 \}$ and consider the unsigned Laplacian $L^+$ on the same graph.  By Cheeger's inequality there must be a subset $S \subseteq V$ with $\Phi(S) \leq \sqrt{2 D_{\max}  \lambda^+_{2}}$, and for any subset $S$ with expansion $\Phi(S)$ satisfying this relation we have }
$$
\lambda^\sigma_2 \leq 2\sqrt{2 D_{\max}}\left( \frac{\vol(\Omega)}{\vol(S \cap \Omega)}\sqrt{\lambda^{\sigma}_1} + \frac{2 \; \vol(S)}{\vol(S \cap \Omega)}\sqrt{\lambda^+_{2}}  \right)
$$
\end{theorem}
The proof is presented in Appendix~\ref{app:sgt}, with the proof strategy illustrated in Fig.~\ref{fig:UnsignedHigherCheeger}.
However, using simpler techniques, we can also obtain the following theorem upper bounding $\lambda_2^+$.
\begin{theorem}(converse bounds): For any graph $G$ and any signature $\sigma$, we have
$$
\lambda_2^+ \leq \sqrt{2 D_{\max} \lambda^\sigma_2}
$$
\end{theorem}
In the context of $n$-qubit local Hamiltonians with bounded interaction degree one always has $D_{\max} = \mathcal{O}(n)$.  The fact that we have $\sqrt{\lambda^\sigma_1}$ and $\sqrt{\lambda_2^+}$ on the RHS weakens the bound, but in appendix~\ref{app:sgt} we describe how improved Cheeger inequalities and standard assumptions about the low-energy spectra of QAO Hamiltonians (in particular, that their exists a gap somewhere in the low energy spectrum) would eliminate these square roots and strengthen the bound.

Finally, recall that $\Omega$ is the size of the basis support set of the non-stoquastic Hamiltonian, and $S$ is the size of the set that minimizes the bottleneck ratio in the ground state of the stoquastic Hamiltonian.  Typically, $S$ is large and encompasses a constant fraction of the total number of vertices (for example, in a path graph with $N$ vertices $\vol(S) = N/2$, and similarly for a hypercube with $N$ vertices).  Therefore in many cases this ratio of volume factors is $\mathcal{O}(1)$, and together with the improved Cheeger inequality the bound becomes as tight as we expect.  In contrast, there will also be some fraction of cases in which $\Omega$ or the intersection $\Omega \cap S$ is very small, and in these cases the bound predicts that the non-stoquastic first excited energy can be much larger than its stoquastic counterpart.  These correspond to cases in which the stoquastic and non-stoquastic ground states are very different from one another.

In conclusion, if larger subsets tend to have (proportionally) larger phase frustration, then the stoquastic and non-stoquastic spectra are similar with generally larger gaps between the stoquastic eigenvalues.  On the other hand, if the phase frustration is nonuniformly distributed then the non-stoquastic eigenstates can look very different from the stoquastic ones.  Specifically, the subsets with small expansion will have extra large frustration, so that the sets that simultaneously minimize expansion and frustration look very different from the ones that minimize the expansion by itself.  For these cases the stoquastic and non-stoquastic spectra can look almost arbitrarily different from one another.

\subsection{Important caveat} \label{sec:Caveat}

Our results so far have concerned the spectrum of the Hamiltonian, but adiabatic evolutions are sometimes restricted to specific subspaces of the entire Hilbert space due to symmetries of the Hamiltonian.  We will refer to this as the `evolution subspace.' In this case, the relevant minimum gap for the adiabatic condition is not the energy gap between the ground state and first excited state of the total Hamiltonian but the energy gap between the lowest two energy states in evolution subspace, which need not necessarily coincide with the ground state and first excited state of the Hamiltonian.  In this case, it may very well be that the non-stoquastic Hamiltonian gap is smaller than its stoquastic counterpart, with the opposite being the case in the evolution subspace.

\section{Numerical simulations}

To complete our discussion, in this section we provide numerical evidence in support of the analytical analysis presented in the preceding sections. We demonstrate that the stoquasticized variants of various classes of random matrices become increasingly more favorable to their non-stoquastic versions with increasing system size. We study the gaps of dense and local matrices as well as quantum annealing runtimes for a few classes of problems. In each case we consider both the de-signed and shifted stoquastizations. 

\subsection{Dense matrices} \label{sec:densematrices}

\begin{figure}[t]
\begin{center}
\includegraphics[width=.9\columnwidth]{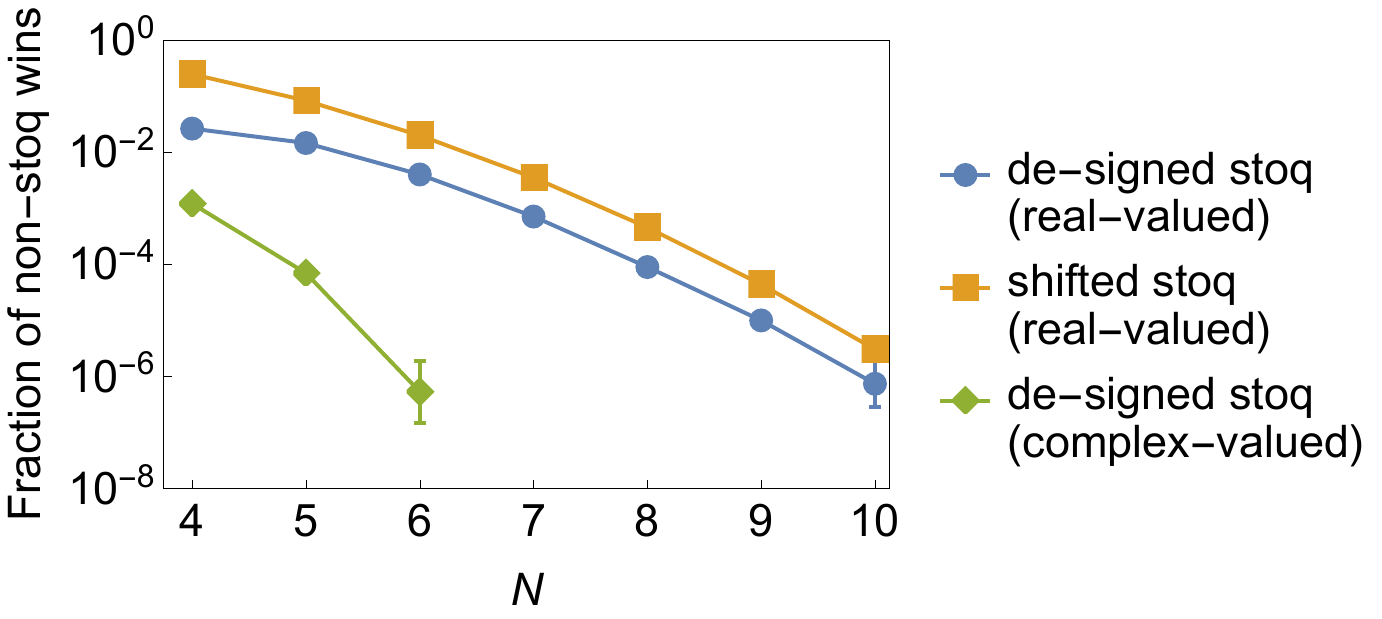}
\caption{Fraction of non-stoquastic wins for dense matrices (described in Sec.~\ref{sec:densematrices}) as a function of matrix dimension $N$. Each data point is the mean over $10^7$ matrix instances, with the error bars being twice the standard error of the mean.}
\label{fig:denseScaling}
\end{center}
\end{figure}

We start off by considering random dense matrices. Here, we generate random (i) real-valued and (ii) complex-valued Hermitian matrices of different dimensions and compute their gaps. In the real case, matrix entries are drawn independently and uniformly from the range $[-1,1]$ whereas in the complex case the sampling is carried out for the real and imaginary parts separately. To enforce Hermiticity, we average the generated matrices with their conjugate transpositions.  As a next step, we calculate numerically the gaps of the stoquasticized counterparts of the generated matrices. 
The figure of merit we focus on is the `fraction of non-stoquastic wins'--- the fraction of occurrences for which the energy gap of the (generally) non-stoquastic matrix is strictly larger than its stoquasticized counterpart. We examine the behavior of this fraction as a function of matrix dimension. The results are summarized in Fig.~\ref{fig:denseScaling}. As is evident, the fraction decays with the matrix dimension, and already at $N=10$ the fraction is incredibly small, which is consistent with our analytical derivation in Sec.~\ref{sec:RandomMatrices}. 

\subsection{Minimum gaps Max-Cut}

When written in terms of Pauli operators, dense matrices are highly non-local in nature and as such are not expected to be easily realizable  in a physical implementation of QAO.  We therefore also consider the more physical setting where the Hamiltonian has bounded locality.
Furthermore, the efficiency of QAO is determined by the minimum gap along the the entire interpolation from the initial to the final Hamiltonian.  To address these two issues, 
 we consider a time-dependent Hamiltonian of the form~\cite{nonstoq3}:
 \beq \label{eqt:H}
 H(s)=(1-s)H_D+s(1-s) H_C + s H_I \,.
 \eeq
where $s$ is the dimensionless annealing parameter, $H_D$ is the initial `driver' Hamiltonian, $H_I$ is the final Hamiltonian that encodes the computational problem, and $H_C$ is a catalyst Hamiltonian.  We take $H_D$ to be the standard transverse-field Hamiltonian, $H_D=-\sum_i X_i$. We take $H_I$ to be an Ising Hamiltonian representing Max-Cut problem instances defined on random 3-regular graphs  \hbox{$H_I=\sum_{\langle i j \rangle} Z_i Z_j$}~\cite{farhi:12,ODE}. This class of problem instances is known to be NP-hard (see, e.g., Refs.~\cite{farhi:12,Liu:2015}).  The Ising Hamiltonian has each spin coupled antiferromagnetically (with strength $J_{ij} =1$) with exactly three other spins picked at random. To incorporate  non-stoquasticity~\cite{nonstoq1,nonstoq2,PhysRevB.95.184416,PhysRevA.95.042321} we choose the catalyst Hamiltonian $H_C$ to be
 \beq \label{eqt:He}
 H_C= \sum_{\langle i j \rangle} \alpha_{ij} X_i X_j
 \eeq
with  the same connectivities $\langle i j \rangle$ as $H_I$.
The coefficients $\alpha_{ij}$ are chosen: (i) uniformly from $[-1,1]$ and (ii) randomly with equal probability from the set $\{-1,1\}$. For the Hamiltonian $H(s)$ above, the off-diagonal entries arise solely from purely off-diagonal Pauli terms, so we implement de-signed stoquastization by the replacement $ \alpha_{ij} \to -| \alpha_{ij}|$ and shifted stoquastization (Eq.~\eqref{eqt:shift1}) by the replacement $ \alpha_{ij} \to \frac{1}{2} \left( \alpha_{ij} - 1 \right)$.  Our choice of a shift by $1$ for shifted stoquastization is constant for all instances in both classes of problems, even if the individual instances with uniform $\alpha \in [-1,1]$ may exhibit a maximum value of $\alpha_{ij}$ that is less than $1$.  

Before proceeding, we note that as defined, the Hamiltonian Eq.~\eqref{eqt:H} is invariant under the transformation \hbox{$P = \prod_i X_i$}, which we refer to as the global bit-flip symmetry of the Hamiltonian.  Since the ground state of $H(0)$ is the uniform superposition state and has eigenvalue $+1$ under $P$, the evolution under $H(s)$ with this initial state is restricted to the $P = +1$ subspace~\cite{CQA,CQA2}.  This subspace is our evolution subspace (see Sec.~\ref{sec:Caveat}), and we denote its energy eigenvalues by $\varepsilon_i(s)$ for $i=0, 1, \dots, 2^{n-1}-1$.  Therefore, the relevant gap for the QAO algorithm is the minimum gap in this subspace.  This is relevant because the lowest energy state in the evolution subspace may not be correspond to the global ground state, whose eigenvalue we denote by $E_0(s)$.   We will see that this may have important consequences.

We study the minimum gaps of randomly generated $n$-spin problems with sizes ranging from  $n=6$ to $n=20$. For each size we generate 100 random instances with a unique satisfying assignment (up to the global bit-flip symmetry) and further consider up to 100 non-stoquastic $\{ \alpha_{ij}\}$ realizations for each Max-Cut instance yielding a maximum of $10^4$ Hamiltonian realizations per problem size.  

\begin{figure}[t]
\begin{center}
\includegraphics[width=.85\columnwidth]{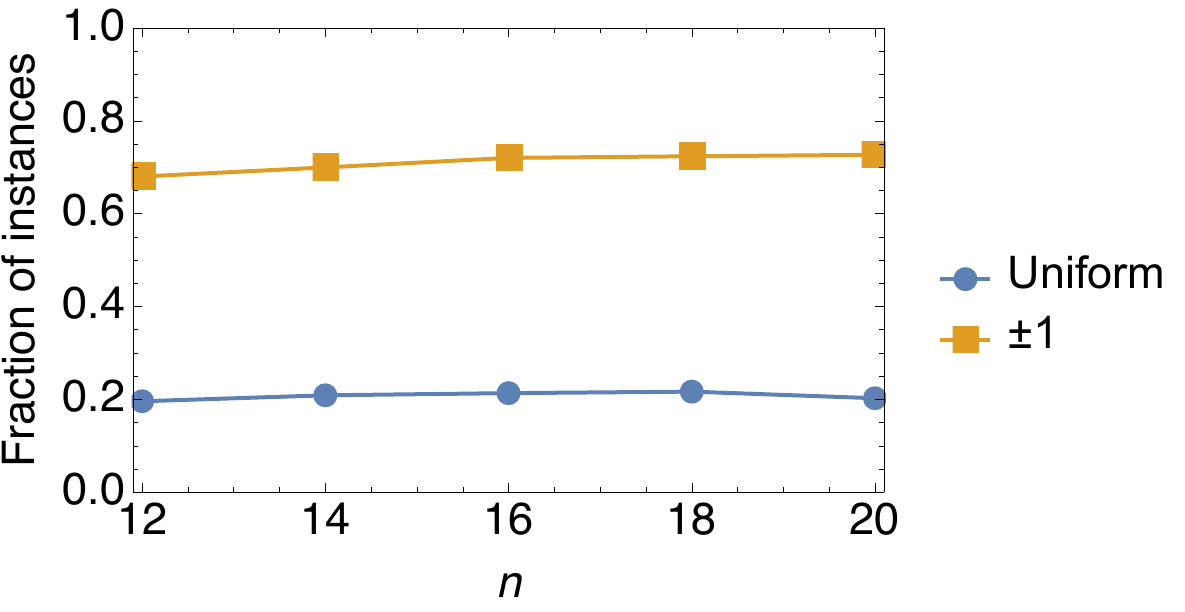}
\caption{Fraction of non-stoquastic instances for Max-Cut problems with an interpolating Hamiltonian of the form in Eq.~\eqref{eqt:H} that have $E_0(s_\ast) < \varepsilon_0(s_\ast)$ at their minimum gap point $s_\ast$. Error bars correspond to $2 \sigma$ confidence calculated with a bootstrap over the instances, $10^4$ instances. Solid lines connecting the data points are to aid the eye.}
\label{fig:fractionNotinP}
\end{center}
\end{figure}

We first show in Fig.~\ref{fig:fractionNotinP} that at the point $s_\ast$ in the interpolation where the minimum gap occurs, a non-negligible fraction of the non-stoquastic instances have $E_0(s_\ast) < \varepsilon_0(s_\ast)$, meaning the lowest energy state in the evolution subspace is not the global ground state.  We therefore choose to distinguish between these two cases, when $E_0(s_\ast) = \varepsilon_0(s_\ast)$ and when $E_0(s_\ast) < \varepsilon_0(s_\ast)$. While the latter fraction appears to be more or less constant over the range of $n$ studied, we cannot rule out the possibility that it will decay to zero at larger $n$ values.   

\begin{figure}[t]
\begin{center}
{\includegraphics[width=.85\columnwidth]{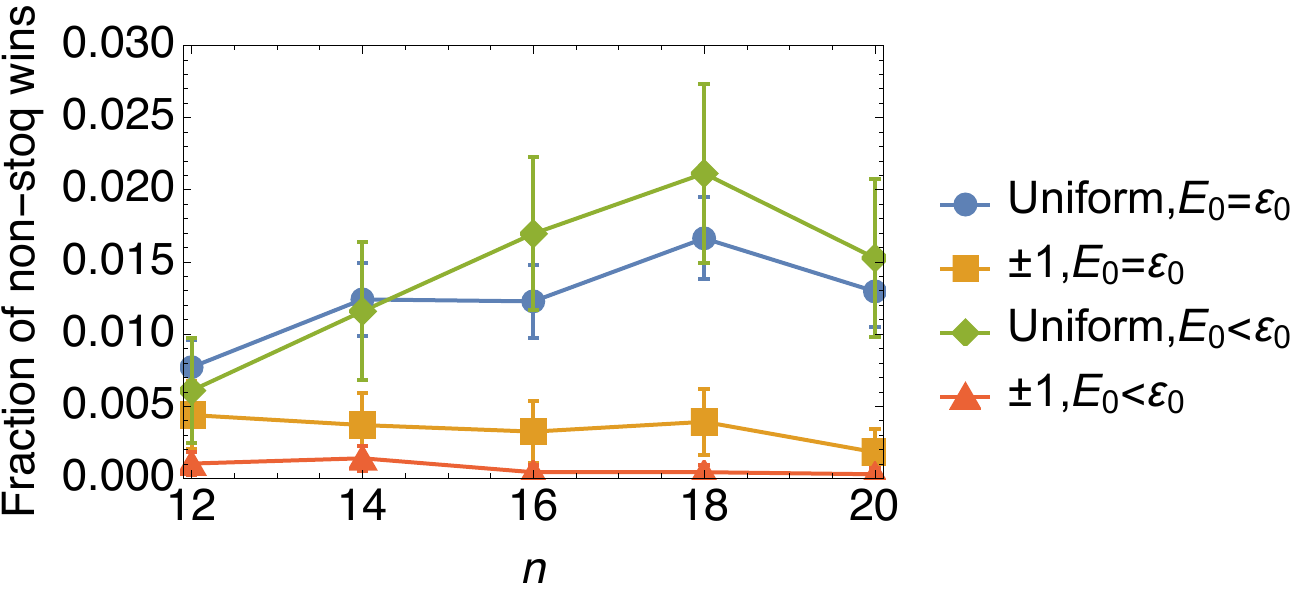}}
{\includegraphics[width=.85\columnwidth]{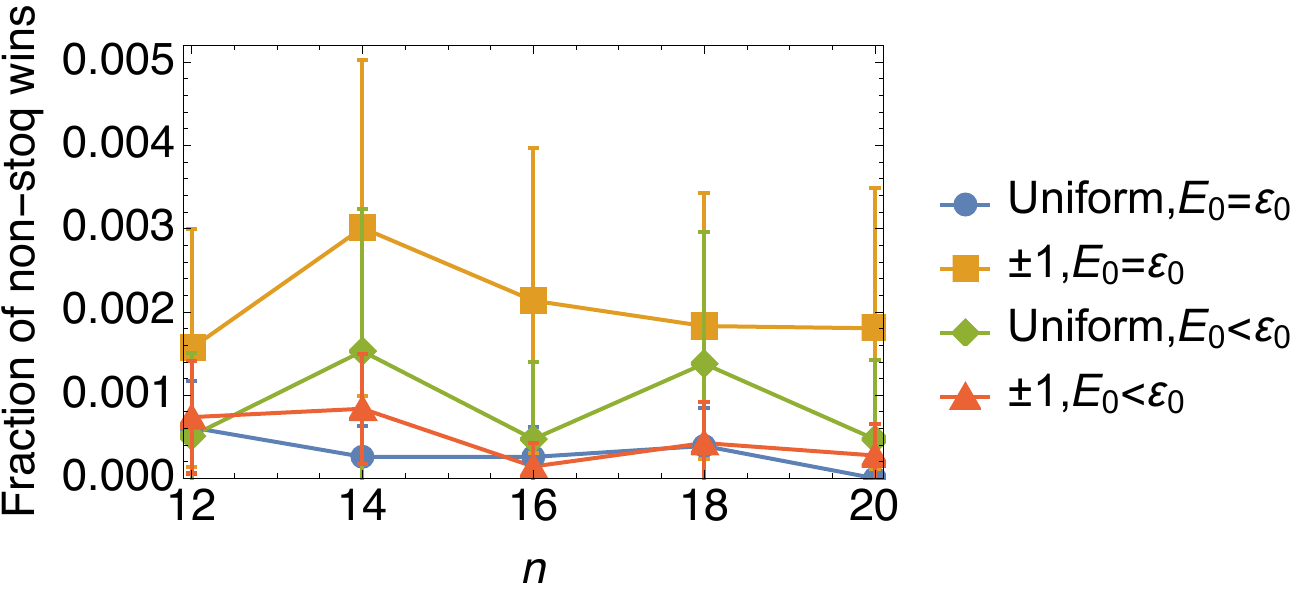}}
\caption{Fraction of non-stoquastic occurrences of larger non-stoquastic minimum gap relative to their stoquasticized analogues, `de-signed' (top) and `shifted' (bottom), as a function of problem size, for Max-Cut problems with an interpolating Hamiltonian of the form in Eq.~\eqref{eqt:H}. Error bars correspond to $2 \sigma$ confidence calculated with a bootstrap over $10^4$ instances.}
\label{fig:2local}
\end{center}
\end{figure}

For every Hamiltonian instance, we inspect the minimum gap of $H(s)$ throughout the evolution and compare it against the minimum gap of its stoquasticized variants. The results are summarized in Fig.~\ref{fig:2local} showing the `fraction of non-stoquastic wins' as a function of system size. As both panels of the figure indicate, the fraction is typically very low, more so for the shifted case where the $2\sigma$ error bars prevent us from concluding anything statistically relevant.  We therefore focus on the de-signed case.  We first observe that that the $\pm1$ non-stoqaustic instances exhibit measurably fewer wins despite having a larger fraction of cases with $E_0(s_\ast) < \epsilon_0(s_\ast)$ (Fig.~\ref{fig:fractionNotinP}).  For the uniform case, we observe that while the instances with $E_0(s_\ast) = \varepsilon_0(s_\ast)$ appear to have a constant fraction of wins as a function of system size, the fraction of wins with instances with $E_0(s_\ast) < \varepsilon_0(s_\ast)$ grows with system size $n$.  While this result may indicate a positive trend, it is important to emphasize that when compared to \emph{both} the de-signed and shifted cases, the results are more in line with the shifted case, meaning that the shifted case almost always beats the non-stoquastic case.

\subsection{Time evolution simulations}

\begin{figure}[t]
\begin{center}
\includegraphics[width=.85\columnwidth]{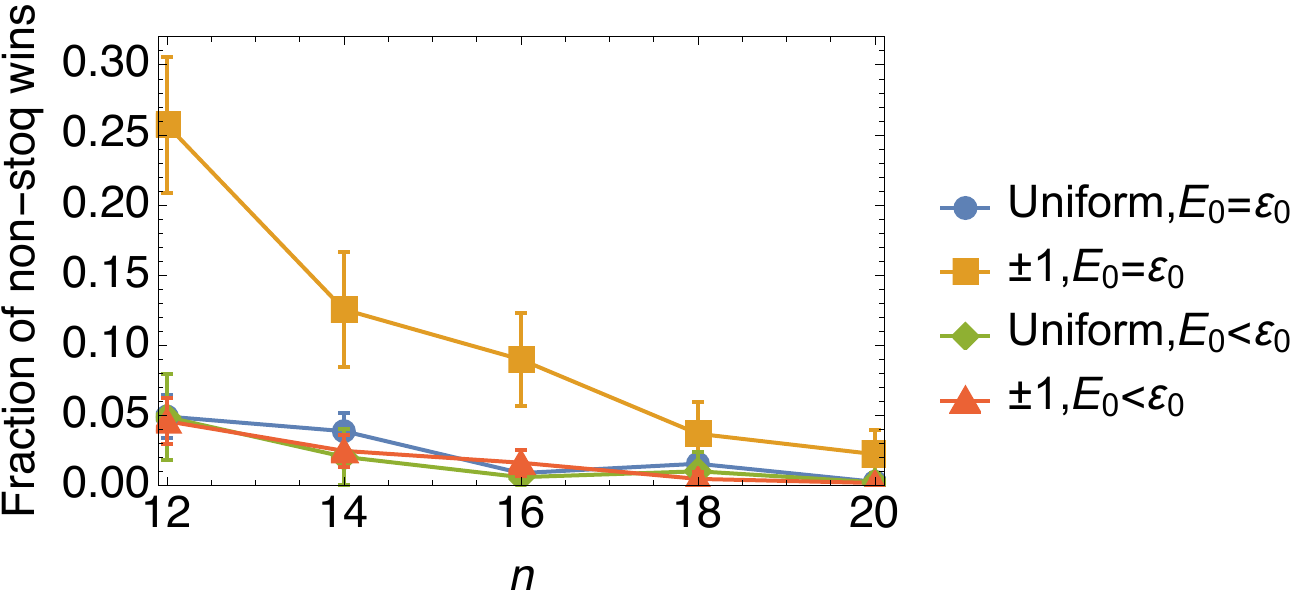}
\includegraphics[width=.85\columnwidth]{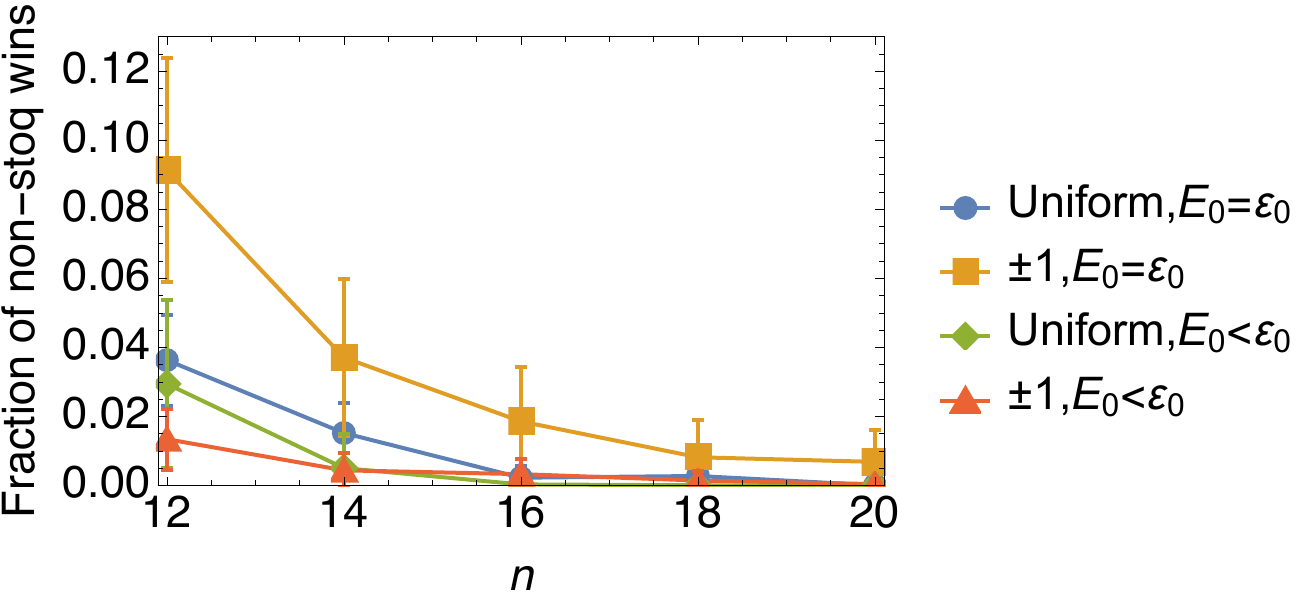}
\caption{Fraction of non-stoquastic occurrences of smaller non-stoquastic average time to solution relative to their stoquasticized analogues as a function of problem size, for Max-Cut problems with an interpolating Hamiltonian of the form in Eq.~\eqref{eqt:H}. Error bars correspond to $2 \sigma$ confidence calculated with a bootstrap over $10^3$ instances.}
\label{fig:timeEvol}
\end{center}
\end{figure}

We note that the fraction of non-stoquastic wins as measured by the gaps (Fig.~\ref{fig:2local}) does not necessarily translate to the same fraction of wins as measured by an optimized computational cost $t_f/p_{\mathrm{GS}}$, often referred to as the `average time to solution'.  For each instance, we find the annealing time that minimizes this quantity and calculate the median computational cost as a function of system size for the same group of instances.  We show our results in Fig.~\ref{fig:timeEvol}, and we see no increase in the fraction of non-stoquastic wins.  This discrepancy between the gaps and the computational cost can be attributed to the smallness of the problem instances: these small instances still have relatively large gaps (we still do not see the exponential decay with system size that might be expected for this class of instances), so the gap alone does not quantitatively predict the computational cost.  

\section{Conclusions and discussion}

In this study, we provided analytical as well as numerical evidence in favor of the assertion that non-stoquastoic driver Hamiltonians are inferior to their stoquastic variants for quantum annealing optimization tasks. We analyzed the gaps of several types of random non-stoquastic Hamiltonians comparing them against their `stoquasticized' counterparts.
We find that generically the non-positivity of the latter Hamiltonians renders their gap larger than that of non-stoquastic ones, making non-stoquastic Hamiltonian less favorable for quantum annealing optimization. 
Our results imply that stoquastic Hamiltonians should generally be preferable to non-stoquastic ones, at least as far as runtimes of quantum adiabatic algorithms are concerned.

 It should be noted that examples to the contrary nonetheless exist. One well understood example where a non-stoquastic Hamiltonian exhibits an exponentially larger minimum gap over its de-signed counterpart  is the case of the ferromagnetic fully-connected $p$-spin models with a non-stoquastic intermediate (or catalyst) Hamiltonian~\cite{Seki:2012,Seoane:2012uq,nonstoq2,PhysRevA.95.042321,Alb2019,Dur2019}, with the caveat that while it is non-stoquastic in the computational $Z$ basis, it is stoquastic in the $X$ basis.  

Another example exhibiting a similar non-stoquastic advantage over its de-signed counterpart is the introduction of a non-stoquastic intermediate Hamiltonian in the strong-weak cluster problem \cite{oneOverF2} between the two clusters \cite{Alb2019,Tak2020}.  This example illustrates an important point: while one choice of intermediate Hamiltonian exhibits an exponentially larger minimum gap over its de-signed counterpart, other choices of the intermediate Hamiltonian can exhibit the opposite effect, with the de-signed Hamiltonian exhibiting an exponentially larger minimum gap over the non-stoquastic Hamiltonian \cite{Tak2020}.  This example illustrates that even if one adiabatic path may exhibit an advantageous minimum gap scaling for the non-stoquastic Hamiltonian, there may be \emph{other} adiabatic paths with stoquastic Hamiltonians that exhibit similar minimum gap scalings. 

While the above examples provide flashes of optimism about   the prospects of non-stoquastic drivers in QAO, our results call into question the promise attributed to non-stoquastic drivers to serve as \emph{generic} catalysts of quantum speedups. 

We note that we have not studied here the possibility of specifically tailoring driver Hamiltonians (stoquastic as well as non-stoquastic) to instances of optimization problems with the goal of raising the minimum gap along the evolution. Since nonetheless such tailoring usually requires detailed knowledge of the spectrum of the problem to be solved, in most cases doing so optimally may turn out to be as difficult to gain knowledge about as the original problem to be solved.

\begin{acknowledgements}
Computation for the work described in this paper was supported by the University of Southern California's Center for High-Performance Computing (hpc.usc.edu) and by ARO grant number W911NF1810227.
The research is based upon work (partially) supported by the Office of
the Director of National Intelligence (ODNI), Intelligence Advanced
Research Projects Activity (IARPA), via the U.S. Army Research Office
contract W911NF-17-C-0050. This material is based on research sponsored by the Air Force Research laboratory under
agreement number FA8750-18-1-0044. The U.S. Government is authorized to reproduce and distribute
reprints for Governmental purposes notwithstanding any copyright notation thereon."
The views and conclusions contained herein are
those of the authors and should not be interpreted as necessarily
representing the official policies or endorsements, either expressed or
implied, of the ODNI, IARPA, or the U.S. Government. The U.S. Government
is authorized to reproduce and distribute reprints for Governmental
purposes notwithstanding any copyright annotation thereon.
\end{acknowledgements}
%\bibliography{refs}

\begin{thebibliography}{72}
\providecommand{\natexlab}[1]{#1}
\providecommand{\url}[1]{\texttt{#1}}
\expandafter\ifx\csname urlstyle\endcsname\relax
  \providecommand{\doi}[1]{doi: #1}\else
  \providecommand{\doi}{doi: \begingroup \urlstyle{rm}\Url}\fi

\bibitem[Aharonov et~al.(2008)Aharonov, van Dam, Kempe, Landau, Lloyd, and
  Regev]{aharonov2008adiabatic}
Dorit Aharonov, Wim van Dam, Julia Kempe, Zeph Landau, Seth Lloyd, and Oded
  Regev.
\newblock Adiabatic quantum computation is equivalent to standard quantum
  computation.
\newblock \emph{SIAM Review}, 50\penalty0 (4):\penalty0 755--787, 2008.
\newblock \doi{https://doi.org/10.1137/080734479}.

\bibitem[{Al-Shimary} and {Pachos}(2010)]{al2010energy}
Abbas {Al-Shimary} and Jiannis~K. {Pachos}.
\newblock {Energy gaps of Hamiltonians from graph Laplacians}.
\newblock \emph{arXiv e-prints}, art. arXiv:1010.4130, October 2010.

\bibitem[Albash(2019)]{Alb2019}
Tameem Albash.
\newblock Role of nonstoquastic catalysts in quantum adiabatic optimization.
\newblock \emph{Phys. Rev. A}, 99:\penalty0 042334, Apr 2019.
\newblock \doi{https://doi.org/10.1103/PhysRevA.99.042334}.

\bibitem[Albash et~al.(2017)Albash, Wagenbreth, and Hen]{ODE}
Tameem Albash, Gene Wagenbreth, and Itay Hen.
\newblock Off-diagonal expansion quantum monte carlo.
\newblock \emph{Phys. Rev. E}, 96:\penalty0 063309, Dec 2017.
\newblock \doi{https://doi.org/10.1103/PhysRevE.96.063309}.

\bibitem[Amin and Choi(2009)]{Ami2009}
M.~H.~S. Amin and V.~Choi.
\newblock First-order quantum phase transition in adiabatic quantum
  computation.
\newblock \emph{Phys. Rev. A}, 80:\penalty0 062326, Dec 2009.
\newblock \doi{https://doi.org/10.1103/PhysRevA.80.062326}.

\bibitem[{Andriyash} and {Amin}(2017)]{And2017}
Evgeny {Andriyash} and Mohammad~H. {Amin}.
\newblock {Can quantum Monte Carlo simulate quantum annealing?}
\newblock \emph{arXiv e-prints}, art. arXiv:1703.09277, March 2017.

\bibitem[Atay and Liu(2020)]{atay2014cheeger}
Fatihcan~M. Atay and Shiping Liu.
\newblock Cheeger constants, structural balance, and spectral clustering
  analysis for signed graphs.
\newblock \emph{Discrete Mathematics}, 343\penalty0 (1):\penalty0 111616, 2020.
\newblock ISSN 0012-365X.
\newblock \doi{https://doi.org/10.1016/j.disc.2019.111616}.

\bibitem[{A.Yu. Kitaev, A.H. Shen, M.N. Vyalyi}(2000)]{Kitaev:book}
{A.Yu. Kitaev, A.H. Shen, M.N. Vyalyi}.
\newblock \emph{Classical and Quantum Computation}, volume~47 of \emph{Graduate
  Studies in Mathematics}.
\newblock American Mathematical Society, Providence, RI, 2000.

\bibitem[Biamonte and Love(2008)]{Biamonte:07}
Jacob~D. Biamonte and Peter~J. Love.
\newblock Realizable hamiltonians for universal adiabatic quantum computers.
\newblock \emph{Phys. Rev. A}, 78:\penalty0 012352, Jul 2008.
\newblock \doi{https://doi.org/10.1103/PhysRevA.78.012352}.

\bibitem[Boixo et~al.(2016)Boixo, Smelyanskiy, Shabani, Isakov, Dykman,
  Denchev, Amin, Smirnov, Mohseni, and Neven]{oneOverF2}
Sergio Boixo, Vadim~N. Smelyanskiy, Alireza Shabani, Sergei~V. Isakov, Mark
  Dykman, Vasil~S. Denchev, Mohammad~H. Amin, Anatoly~Yu Smirnov, Masoud
  Mohseni, and Hartmut Neven.
\newblock Computational multiqubit tunnelling in programmable quantum
  annealers.
\newblock \emph{Nature Communications}, 7:\penalty0 10327, 01 2016.
\newblock \doi{https://doi.org/10.1038/ncomms10327}.

\bibitem[Bravyi and Terhal(2009)]{bravyi2009complexity}
S.~Bravyi and B.~Terhal.
\newblock Complexity of stoquastic frustration-free hamiltonians.
\newblock \emph{SIAM Journal on Computing}, 39\penalty0 (4):\penalty0
  1462--1485, 2015/05/22 2009.
\newblock \doi{https://doi.org/10.1137/08072689X}.

\bibitem[Bravyi(2015)]{bravyi2015monte}
Sergey Bravyi.
\newblock Monte carlo simulation of stoquastic hamiltonians.
\newblock \emph{Quantum Info. Comput.}, 15\penalty0 (13–14):\penalty0
  1122–1140, October 2015.
\newblock ISSN 1533-7146.

\bibitem[Bravyi and Gosset(2017)]{bravyi2017polynomial}
Sergey Bravyi and David Gosset.
\newblock Polynomial-time classical simulation of quantum ferromagnets.
\newblock \emph{Phys. Rev. Lett.}, 119:\penalty0 100503, Sep 2017.
\newblock \doi{https://doi.org/10.1103/PhysRevLett.119.100503}.

\bibitem[{Bravyi} et~al.(2006){Bravyi}, {Bessen}, and
  {Terhal}]{bravyi2006merlin}
Sergey {Bravyi}, Arvid~J. {Bessen}, and Barbara~M. {Terhal}.
\newblock {Merlin-Arthur Games and Stoquastic Complexity}.
\newblock \emph{arXiv e-prints}, art. quant-ph/0611021, November 2006.

\bibitem[Bravyi et~al.(2008)Bravyi, Divincenzo, Oliveira, and
  Terhal]{bravyi2006complexity}
Sergey Bravyi, David~P. Divincenzo, Roberto Oliveira, and Barbara~M. Terhal.
\newblock The complexity of stoquastic local hamiltonian problems.
\newblock \emph{Quantum Info. Comput.}, 8\penalty0 (5):\penalty0 361–385, May
  2008.
\newblock ISSN 1533-7146.

\bibitem[Brooke et~al.(1999)Brooke, Bitko, F., Rosenbaum, and
  Aeppli]{Brooke1999}
J.~Brooke, D.~Bitko, T.~F., Rosenbaum, and G.~Aeppli.
\newblock Quantum annealing of a disordered magnet.
\newblock \emph{Science}, 284\penalty0 (5415):\penalty0 779--781, 1999.
\newblock \doi{https://doi.org/10.1126/science.284.5415.779}.

\bibitem[Chung and Graham(1997)]{chung1997spectral}
Fan~RK Chung and Fan~Chung Graham.
\newblock \emph{Spectral graph theory}.
\newblock Number~92 in CBMS Regional Conference Series. American Mathematical
  Society, 1997.

\bibitem[{Crosson} and {Bowen}(2017)]{crosson2017quantum}
Elizabeth {Crosson} and John {Bowen}.
\newblock {Quantum ground state isoperimetric inequalities for the energy
  spectrum of local Hamiltonians}.
\newblock \emph{arXiv e-prints}, art. arXiv:1703.10133, March 2017.

\bibitem[Crosson and Harrow(2016)]{crosson2016simulated}
Elizabeth Crosson and Aram~W Harrow.
\newblock Simulated quantum annealing can be exponentially faster than
  classical simulated annealing.
\newblock In \emph{2016 IEEE 57th Annual Symposium on Foundations of Computer
  Science (FOCS)}, pages 714--723. IEEE, 2016.
\newblock \doi{https://doi.org/10.1109/FOCS.2016.81}.

\bibitem[{Crosson} and {Harrow}(2018)]{crosson2018rapid}
Elizabeth {Crosson} and Aram~W. {Harrow}.
\newblock {Rapid mixing of path integral Monte Carlo for 1D stoquastic
  Hamiltonians}.
\newblock \emph{arXiv e-prints}, art. arXiv:1812.02144, December 2018.

\bibitem[{Crosson} et~al.(2014){Crosson}, {Farhi}, {Yen-Yu Lin}, {Lin}, and
  {Shor}]{nonstoq3}
Elizabeth {Crosson}, Edward {Farhi}, Cedric {Yen-Yu Lin}, Han-Hsuan {Lin}, and
  Peter {Shor}.
\newblock {Different Strategies for Optimization Using the Quantum Adiabatic
  Algorithm}.
\newblock \emph{arXiv e-prints}, art. arXiv:1401.7320, January 2014.

\bibitem[Durkin(2019)]{Dur2019}
Gabriel~A. Durkin.
\newblock Quantum speedup at zero temperature via coherent catalysis.
\newblock \emph{Phys. Rev. A}, 99:\penalty0 032315, Mar 2019.
\newblock \doi{https://doi.org/10.1103/PhysRevA.99.032315}.

\bibitem[Elgart and Hagedorn(2012)]{Elgart}
Alexander Elgart and George~A. Hagedorn.
\newblock A note on the switching adiabatic theorem.
\newblock \emph{Journal of Mathematical Physics}, 53\penalty0 (10):\penalty0
  102202, 2012.
\newblock \doi{https://doi.org/10.1063/1.4748968}.

\bibitem[{Farhi} et~al.(2000){Farhi}, {Goldstone}, {Gutmann}, and
  {Sipser}]{farhi_quantum_2000}
Edward {Farhi}, Jeffrey {Goldstone}, Sam {Gutmann}, and Michael {Sipser}.
\newblock {Quantum Computation by Adiabatic Evolution}.
\newblock \emph{arXiv e-prints}, art. quant-ph/0001106, January 2000.

\bibitem[{Farhi} et~al.(2002){Farhi}, {Goldstone}, and {Gutmann}]{farhi:02}
Edward {Farhi}, Jeffrey {Goldstone}, and Sam {Gutmann}.
\newblock {Quantum Adiabatic Evolution Algorithms versus Simulated Annealing}.
\newblock \emph{arXiv e-prints}, art. quant-ph/0201031, January 2002.

\bibitem[Farhi et~al.(2012)Farhi, Gosset, Hen, Sandvik, Shor, Young, and
  Zamponi]{farhi:12}
Edward Farhi, David Gosset, Itay Hen, A.~W. Sandvik, Peter Shor, A.~P. Young,
  and Francesco Zamponi.
\newblock Performance of the quantum adiabatic algorithm on random instances of
  two optimization problems on regular hypergraphs.
\newblock \emph{Phys. Rev. A}, 86:\penalty0 052334, Nov 2012.
\newblock \doi{https://doi.org/10.1103/PhysRevA.86.052334}.

\bibitem[F{\'e}ral and P{\'e}ch{\'e}(2007)]{feral2007largest}
Delphine F{\'e}ral and Sandrine P{\'e}ch{\'e}.
\newblock The largest eigenvalue of rank one deformation of large wigner
  matrices.
\newblock \emph{Communications in Mathematical Physics}, 272\penalty0
  (1):\penalty0 185--228, 2007.
\newblock \doi{https://doi.org/10.1007/s00220-007-0209-3}.

\bibitem[Finnila et~al.(1994)Finnila, Gomez, Sebenik, Stenson, and
  Doll]{finnila_quantum_1994}
A.~B. Finnila, M.~A. Gomez, C.~Sebenik, C.~Stenson, and J.~D. Doll.
\newblock Quantum annealing: A new method for minimizing multidimensional
  functions.
\newblock \emph{Chemical Physics Letters}, 219\penalty0 (5--6):\penalty0
  343--348, 3 1994.
\newblock \doi{https://doi.org/10.1016/0009-2614(94)00117-0}.

\bibitem[Franklin(1993)]{weyl}
Joel.~N Franklin.
\newblock \emph{Matrix Theory (Dover Books on Mathematics)}.
\newblock Dover Publications, Inc., Mineola, NY, USA, 1993.

\bibitem[Gournay(2016)]{gournay2016isoperimetric}
Antoine Gournay.
\newblock An isoperimetric constant for signed graphs.
\newblock \emph{Expositiones Mathematicae}, 34\penalty0 (3):\penalty0 339 --
  351, 2016.
\newblock ISSN 0723-0869.
\newblock \doi{https://doi.org/10.1016/j.exmath.2015.10.002}.

\bibitem[Gupta and Hen(2020)]{gupta2019}
Lalit Gupta and Itay Hen.
\newblock Elucidating the interplay between non-stoquasticity and the sign
  problem.
\newblock \emph{Advanced Quantum Technologies}, 3\penalty0 (1):\penalty0
  1900108, 2020.
\newblock \doi{https://doi.org/10.1002/qute.201900108}.

\bibitem[{Harrow} et~al.(2019){Harrow}, {Mehraban}, and
  {Soleimanifar}]{harrow2019classical}
Aram {Harrow}, Saeed {Mehraban}, and Mehdi {Soleimanifar}.
\newblock {Classical algorithms, correlation decay, and complex zeros of
  partition functions of quantum many-body systems}.
\newblock \emph{arXiv e-prints}, art. arXiv:1910.09071, October 2019.

\bibitem[Hastings(2013)]{Hastings:2013kk}
Matthew~B. Hastings.
\newblock Obstructions to classically simulating the quantum adiabatic
  algorithm.
\newblock \emph{Quantum Info. Comput.}, 13\penalty0 (11–12):\penalty0
  1038–1076, November 2013.
\newblock ISSN 1533-7146.

\bibitem[Hen(2012)]{hen:12}
Itay Hen.
\newblock Excitation gap from optimized correlation functions in quantum monte
  carlo simulations.
\newblock \emph{Phys. Rev. E}, 85:\penalty0 036705, Mar 2012.
\newblock \doi{https://doi.org/10.1103/PhysRevE.85.036705}.

\bibitem[Hen and Sarandy(2016)]{CQA2}
Itay Hen and Marcelo~S. Sarandy.
\newblock Driver hamiltonians for constrained optimization in quantum
  annealing.
\newblock \emph{Phys. Rev. A}, 93:\penalty0 062312, Jun 2016.
\newblock \doi{https://doi.org/10.1103/PhysRevA.93.062312}.

\bibitem[Hen and Spedalieri(2016)]{CQA}
Itay Hen and Federico~M. Spedalieri.
\newblock Quantum annealing for constrained optimization.
\newblock \emph{Phys. Rev. Applied}, 5:\penalty0 034007, Mar 2016.
\newblock \doi{https://doi.org/10.1103/PhysRevApplied.5.034007}.

\bibitem[Hen and Young(2011)]{hen:11}
Itay Hen and A.~P. Young.
\newblock Exponential complexity of the quantum adiabatic algorithm for certain
  satisfiability problems.
\newblock \emph{Phys. Rev. E}, 84:\penalty0 061152, Dec 2011.
\newblock \doi{https://doi.org/10.1103/PhysRevE.84.061152}.

\bibitem[Hormozi et~al.(2017)Hormozi, Brown, Carleo, and
  Troyer]{PhysRevB.95.184416}
Layla Hormozi, Ethan~W. Brown, Giuseppe Carleo, and Matthias Troyer.
\newblock Nonstoquastic hamiltonians and quantum annealing of an ising spin
  glass.
\newblock \emph{Phys. Rev. B}, 95:\penalty0 184416, May 2017.
\newblock \doi{https://doi.org/10.1103/PhysRevB.95.184416}.

\bibitem[Horn et~al.(1990)Horn, Horn, and Johnson]{horn1990matrix}
Roger~A Horn, Roger~A Horn, and Charles~R Johnson.
\newblock \emph{Matrix analysis}.
\newblock Cambridge university press, 1990.

\bibitem[Isakov et~al.(2016)Isakov, Mazzola, Smelyanskiy, Jiang, Boixo, Neven,
  and Troyer]{Isa2016}
Sergei~V. Isakov, Guglielmo Mazzola, Vadim~N. Smelyanskiy, Zhang Jiang, Sergio
  Boixo, Hartmut Neven, and Matthias Troyer.
\newblock Understanding quantum tunneling through quantum monte carlo
  simulations.
\newblock \emph{Phys. Rev. Lett.}, 117:\penalty0 180402, Oct 2016.
\newblock \doi{https://doi.org/10.1103/PhysRevLett.117.180402}.

\bibitem[Jansen et~al.(2007)Jansen, Ruskai, and Seiler]{Jansen:07}
Sabine Jansen, Mary-Beth Ruskai, and Ruedi Seiler.
\newblock Bounds for the adiabatic approximation with applications to quantum
  computation.
\newblock \emph{J. Math. Phys.}, 48\penalty0 (10):\penalty0 --, 2007.
\newblock \doi{https://doi.org/10.1063/1.2798382}.

\bibitem[{Jarret}(2018)]{jarret2018hamiltonian}
Michael {Jarret}.
\newblock {Hamiltonian surgery: Cheeger-type gap inequalities for nonpositive
  (stoquastic), real, and Hermitian matrices}.
\newblock \emph{arXiv e-prints}, art. arXiv:1804.06857, April 2018.

\bibitem[Jarret and Jordan(2015)]{jarret2014adiabatic}
Michael Jarret and Stephen~P. Jordan.
\newblock Adiabatic optimization without local minima.
\newblock \emph{Quantum Info. Comput.}, 15\penalty0 (3–4):\penalty0
  181–199, March 2015.
\newblock ISSN 1533-7146.

\bibitem[Jarret et~al.(2016)Jarret, Jordan, and Lackey]{PhysRevA.94.042318}
Michael Jarret, Stephen~P. Jordan, and Brad Lackey.
\newblock Adiabatic optimization versus diffusion monte carlo methods.
\newblock \emph{Phys. Rev. A}, 94:\penalty0 042318, Oct 2016.
\newblock \doi{https://doi.org/10.1103/PhysRevA.94.042318}.

\bibitem[{Jarret} et~al.(2018){Jarret}, {Lackey}, {Liu}, and
  {Wan}]{jarret2018quantum}
Michael {Jarret}, Brad {Lackey}, Aike {Liu}, and Kianna {Wan}.
\newblock {Quantum adiabatic optimization without heuristics}.
\newblock \emph{arXiv e-prints}, art. arXiv:1810.04686, October 2018.

\bibitem[Kadowaki and Nishimori(1998)]{kadowaki_quantum_1998}
Tadashi Kadowaki and Hidetoshi Nishimori.
\newblock Quantum annealing in the transverse ising model.
\newblock \emph{Phys. Rev. E}, 58:\penalty0 5355--5363, Nov 1998.
\newblock \doi{https://doi.org/10.1103/PhysRevE.58.5355}.

\bibitem[Kato(1950)]{Kato:50}
T.~Kato.
\newblock On the adiabatic theorem of quantum mechanics.
\newblock \emph{J. Phys. Soc. Jap.}, 5:\penalty0 435, 1950.
\newblock \doi{https://doi.org/10.1143/JPSJ.5.435}.

\bibitem[Klassen and Terhal(2019)]{Klassen2019twolocalqubit}
Joel Klassen and Barbara~M. Terhal.
\newblock Two-local qubit {H}amiltonians: when are they stoquastic?
\newblock \emph{{Quantum}}, 3:\penalty0 139, May 2019.
\newblock ISSN 2521-327X.
\newblock \doi{https://doi.org/10.22331/q-2019-05-06-139}.

\bibitem[{Klassen} et~al.(2019){Klassen}, {Marvian}, {Piddock}, {Ioannou},
  {Hen}, and {Terhal}]{2019arXiv190608800K}
Joel {Klassen}, Milad {Marvian}, Stephen {Piddock}, Marios {Ioannou}, Itay
  {Hen}, and Barbara {Terhal}.
\newblock {Hardness and Ease of Curing the Sign Problem for Two-Local Qubit
  Hamiltonians}.
\newblock \emph{arXiv e-prints}, art. arXiv:1906.08800, Jun 2019.

\bibitem[Lidar et~al.(2009)Lidar, Rezakhani, and Hamma]{lidarGap}
Daniel~A. Lidar, Ali~T. Rezakhani, and Alioscia Hamma.
\newblock Adiabatic approximation with exponential accuracy for many-body
  systems and quantum computation.
\newblock \emph{Journal of Mathematical Physics}, 50\penalty0 (10):\penalty0
  102106, 2009.
\newblock \doi{https://doi.org/10.1063/1.3236685}.

\bibitem[Liu et~al.(2015)Liu, Polkovnikov, and Sandvik]{Liu:2015}
Cheng-Wei Liu, Anatoli Polkovnikov, and Anders~W. Sandvik.
\newblock Quantum versus classical annealing: Insights from scaling theory and
  results for spin glasses on 3-regular graphs.
\newblock \emph{Phys. Rev. Lett.}, 114:\penalty0 147203, Apr 2015.
\newblock \doi{https://doi.org/10.1103/PhysRevLett.114.147203}.

\bibitem[Loh et~al.(1990)Loh, Gubernatis, Scalettar, White, Scalapino, and
  Sugar]{Loh-PRB-90}
E.~Y. Loh, J.~E. Gubernatis, R.~T. Scalettar, S.~R. White, D.~J. Scalapino, and
  R.~L. Sugar.
\newblock Sign problem in the numerical simulation of many-electron systems.
\newblock \emph{Phys. Rev. B}, 41:\penalty0 9301--9307, May 1990.
\newblock \doi{https://doi.org/10.1103/PhysRevB.41.9301}.

\bibitem[Martin(2017)]{martin2017frustration}
Florian Martin.
\newblock Frustration and isoperimetric inequalities for signed graphs.
\newblock \emph{Discrete Applied Mathematics}, 217:\penalty0 276 -- 285, 2017.
\newblock ISSN 0166-218X.
\newblock \doi{https://doi.org/10.1016/j.dam.2016.09.015}.

\bibitem[Marvian et~al.(2019)Marvian, Lidar, and Hen]{marvian:2018}
Milad Marvian, Daniel~A. Lidar, and Itay Hen.
\newblock On the computational complexity of curing non-stoquastic
  hamiltonians.
\newblock \emph{Nature Communications}, 10\penalty0 (1):\penalty0 1571, 2019.
\newblock \doi{https://doi.org/10.1038/s41467-019-09501-6}.

\bibitem[Nie et~al.(2013)Nie, Katsura, and Oshikawa]{PhysRevLett.111.100402}
Wenxing Nie, Hosho Katsura, and Masaki Oshikawa.
\newblock Ground-state energies of spinless free fermions and hard-core bosons.
\newblock \emph{Phys. Rev. Lett.}, 111:\penalty0 100402, Sep 2013.
\newblock \doi{https://doi.org/10.1103/PhysRevLett.111.100402}.

\bibitem[Nie et~al.(2018)Nie, Katsura, and Oshikawa]{PhysRevB.97.125153}
Wenxing Nie, Hosho Katsura, and Masaki Oshikawa.
\newblock Particle statistics, frustration, and ground-state energy.
\newblock \emph{Phys. Rev. B}, 97:\penalty0 125153, Mar 2018.
\newblock \doi{https://doi.org/10.1103/PhysRevB.97.125153}.

\bibitem[Nishimori and Takada(2017)]{nonstoq2}
Hideyoshi Nishimori and Kabuki Takada.
\newblock Exponential enhancement of the efficiency of quantum annealing by
  non-stoquastic hamiltonians.
\newblock \emph{Frontiers in ICT}, 4:\penalty0 2, 2017.
\newblock ISSN 2297-198X.
\newblock \doi{https://doi.org/10.3389/fict.2017.00002}.

\bibitem[\"Ozg\"uler et~al.(2018)\"Ozg\"uler, Joynt, and
  Vavilov]{PhysRevA.98.062311}
A.~Bar\i\c{s} \"Ozg\"uler, Robert Joynt, and Maxim~G. Vavilov.
\newblock Steering random spin systems to speed up the quantum adiabatic
  algorithm.
\newblock \emph{Phys. Rev. A}, 98:\penalty0 062311, Dec 2018.
\newblock \doi{https://doi.org/10.1103/PhysRevA.98.062311}.

\bibitem[Roland and Cerf(2002)]{Roland:2002ul}
J{\'e}r{\'e}mie Roland and Nicolas~J. Cerf.
\newblock Quantum search by local adiabatic evolution.
\newblock \emph{Phys. Rev. A}, 65\penalty0 (4):\penalty0 042308--, 03 2002.
\newblock \doi{https://doi.org/10.1103/PhysRevA.65.042308}.

\bibitem[R{\o}nnow et~al.(2014)R{\o}nnow, Wang, Job, Boixo, Isakov, Wecker,
  Martinis, Lidar, and Troyer]{speedup}
Troels~F. R{\o}nnow, Zhihui Wang, Joshua Job, Sergio Boixo, Sergei~V. Isakov,
  David Wecker, John~M. Martinis, Daniel~A. Lidar, and Matthias Troyer.
\newblock Defining and detecting quantum speedup.
\newblock \emph{Science}, 345\penalty0 (6195):\penalty0 420--424, 07 2014.
\newblock \doi{https://doi.org/10.1126/science.1252319}.

\bibitem[Santoro et~al.(2002)Santoro, Marto\v{n}\'{a}k, Tosatti, and
  Car]{Santoro}
Giuseppe~E. Santoro, Roman Marto\v{n}\'{a}k, Erio Tosatti, and Roberto Car.
\newblock Theory of quantum annealing of an {I}sing spin glass.
\newblock \emph{Science}, 295\penalty0 (5564):\penalty0 2427--2430, 2002.
\newblock \doi{https://doi.org/10.1126/science.1068774}.

\bibitem[Seki and Nishimori(2012)]{Seki:2012}
Yuya Seki and Hidetoshi Nishimori.
\newblock Quantum annealing with antiferromagnetic fluctuations.
\newblock \emph{Phys. Rev. E}, 85:\penalty0 051112, May 2012.
\newblock \doi{https://doi.org/10.1103/PhysRevE.85.051112}.

\bibitem[Seoane and Nishimori(2012)]{Seoane:2012uq}
Beatriz Seoane and Hidetoshi Nishimori.
\newblock Many-body transverse interactions in the quantum annealing of the p
  -spin ferromagnet.
\newblock \emph{J. Phys. A}, 45\penalty0 (43):\penalty0 435301, 2012.
\newblock \doi{https://doi.org/10.1088/1751-8113/45/43/435301}.

\bibitem[Susa et~al.(2017)Susa, Jadebeck, and Nishimori]{PhysRevA.95.042321}
Yuki Susa, Johann~F. Jadebeck, and Hidetoshi Nishimori.
\newblock Relation between quantum fluctuations and the performance enhancement
  of quantum annealing in a nonstoquastic hamiltonian.
\newblock \emph{Phys. Rev. A}, 95:\penalty0 042321, Apr 2017.
\newblock \doi{https://doi.org/10.1103/PhysRevA.95.042321}.

\bibitem[Takada et~al.(2020)Takada, Yamashiro, and Nishimori]{Tak2020}
Kabuki Takada, Yu~Yamashiro, and Hidetoshi Nishimori.
\newblock Mean-field solution of the weak-strong cluster problem for quantum
  annealing with stoquastic and non-stoquastic catalysts.
\newblock \emph{Journal of the Physical Society of Japan}, 89\penalty0
  (4):\penalty0 044001, 2020.
\newblock \doi{https://doi.org/10.7566/JPSJ.89.044001}.

\bibitem[Tao(2012)]{tao2012topics}
Terence Tao.
\newblock Topics in random matrix theory.
\newblock \emph{Graduate studies in Mathematics}, 132:\penalty0 46--47, 2012.

\bibitem[Troyer and Wiese(2005)]{Wiese-PRL-05}
Matthias Troyer and Uwe-Jens Wiese.
\newblock Computational complexity and fundamental limitations to fermionic
  quantum monte carlo simulations.
\newblock \emph{Phys. Rev. Lett.}, 94:\penalty0 170201, May 2005.
\newblock \doi{https://doi.org/10.1103/PhysRevLett.94.170201}.

\bibitem[Vinci and Lidar(2017)]{nonstoq1}
Walter Vinci and Daniel~A. Lidar.
\newblock Non-stoquastic hamiltonians in quantum annealing via geometric
  phases.
\newblock \emph{npj Quantum Information}, 3\penalty0 (1):\penalty0 38, 2017.
\newblock \doi{https://doi.org/10.1038/s41534-017-0037-z}.

\bibitem[Wannier(1965)]{wannier:65}
Gregory~H. Wannier.
\newblock Probability of violation of the ehrenfest principle in fast passage.
\newblock \emph{Physics Physique Fizika}, 1:\penalty0 251--253, May 1965.
\newblock \doi{https://doi.org/10.1103/PhysicsPhysiqueFizika.1.251}.

\bibitem[Wigner(1958)]{wigner1958distribution}
Eugene~P. Wigner.
\newblock On the distribution of the roots of certain symmetric matrices.
\newblock \emph{Annals of Mathematics}, 67\penalty0 (2):\penalty0 325--327,
  1958.
\newblock ISSN 0003486X.
\newblock \doi{https://doi.org/10.2307/1970008}.

\bibitem[Young et~al.(2008)Young, Knysh, and Smelyanskiy]{Young2008}
A.~P. Young, S.~Knysh, and V.~N. Smelyanskiy.
\newblock Size dependence of the minimum excitation gap in the quantum
  adiabatic algorithm.
\newblock \emph{Phys. Rev. Lett.}, 101:\penalty0 170503, Oct 2008.
\newblock \doi{https://doi.org/10.1103/PhysRevLett.101.170503}.

\bibitem[Young et~al.(2010)Young, Knysh, and Smelyanskiy]{Young:2010}
A.~P. Young, S.~Knysh, and V.~N. Smelyanskiy.
\newblock First-order phase transition in the quantum adiabatic algorithm.
\newblock \emph{Phys. Rev. Lett.}, 104:\penalty0 020502, Jan 2010.
\newblock \doi{https://doi.org/10.1103/PhysRevLett.104.020502}.

\end{thebibliography}

\appendix

\section{Method for de-signing off-diagonal Pauli terms} \label{app:designing}
Suppose the Hamiltonian has the form $H = H_Z + H_O$, where $H_Z$ is diagonal in the computational basis and $H_O$ is a sum of $k$-local Pauli strings composed of only Pauli $X$ and $Y$ operators.  If $b, b'$ are bit strings that are more than Hamming distance $k$ apart, then $ \bra{b} H \ket{b'} = 0$.  Therefore we can efficiently compute each non-zero off-diagonal element of $H_O$ by the following algorithm.  For each subset of $k$ qubits, one sums together the terms of $H_O$ that act entirely within this subset.  This sum of terms $H_O^{(S)}$ is a Hamiltonian matrix of dimension $2^k \times 2^k$ with $2^k$ non-zero terms, so the corresponding de-signed Hamiltonian $(H_O^{(S)})^\ds$ can be computed in time $\mathcal{O}(2^k)$.  The full de-signed Hamiltonian has the form:
\bea
H^\ds &=& H_Z + H_O^\ds \nonumber \\
&=&  H_Z + \sum_{\substack{S \subseteq \{1,...,n\} \\ |S| = k}} (H_O^{(S)})^\ds \ .
\eea
A Pauli-basis representation can then be computed in the usual way using the Hilbert-Schmidt inner product.  For example, the coefficient of the Pauli $X$ operator acting on the $i$-th qubit in the expansion would be proportional to $\Tr(H_O^\ds X_i)$.

\section{Ground state energies} \label{app:variationalGroundEnergy}
In this section we give the proof of Observation~(1) in the introduction.  Let $H$ have ground state energy $E_0$ and ground state wave function
$$
|\psi\rangle = \sum_{z \in \mathcal{B}} \psi_z | z\rangle
$$
expressed in the basis $\mathcal{B}$.  Therefore we have
\begin{equation}
E = \langle \psi | H | \psi\rangle = \sum_{z \in \mathcal{B}} |\psi_z|^2 H_{zz} + \sum_{\substack{z, z' \in \mathcal{B}\\z\neq z'}} \psi^*_z H_{z,z'} \psi_{z'}
\end{equation} 
Let $H^\ds$ be the de-signed stoquastization of $H$ in the basis $\mathcal{B}$, and consider the following variational ansatz for the ground state of $H^\ds$,
$$
|\phi\rangle = \sum_{z \in \mathcal{B}} |\psi_z| |z\rangle.
$$
The energy of this ansatz is
\begin{align*}
\langle \phi | H^\ds | \phi \rangle &= \sum_{z \in \mathcal{B}} |\psi_z|^2 H_{zz} - \sum_{\substack{z, z' \in \mathcal{B}\\ z\neq z'}} |\psi_z| |H_{z,z'}| |\psi_{z'}|\\
&\leq E
\end{align*}
where the inequality follows because 
$$
\sum_{\substack{z, z' \in \mathcal{B}\\z\neq z'}} \psi^*_z H_{z,z'} \psi_{z'} \leq \sum_{\substack{z, z' \in \mathcal{B}\\ z\neq z'}} |\psi_z| |H_{z,z'}| |\psi_{z'}|
$$
by the triangle inequality for real numbers.  For equality to hold, we must have
$$
\sum_{\substack{z, z' \in \mathcal{B}\\z\neq z'}} \psi^*_z H_{z,z'} \psi_{z'} = \sum_{\substack{z, z' \in \mathcal{B}\\ z\neq z'}} |\psi_z| |H_{z,z'}| |\psi_{z'}|
$$
which can only hold if $\psi^*_z H_{z,z'} \psi_{z'}  = |\psi_z| |H_{z,z'}| |\psi_{z'}|$ for each $z,z' \in \mathcal{B}$.  But this latest condition implies that the following signature matrix,
$$
\Lambda = \sum_{\substack{z \in \mathcal{B} \\ \psi_z \neq 0}} \frac{\psi_{z}}{|\psi_{z}|} |z\rangle \langle z|
$$
satisfies $\Lambda H \Lambda^\dagger = H^\ds$.  Since $\Lambda$ is unitary this implies that $H$ was secretly stoquastic.  

\section{Spectra of signed graphs}\label{app:sgt}

A signed graph $\Gamma = (G,\sigma)$ is a graph $G = (\cV,\cE)$ with vertex set $\cV$ and edge set $\cE$, together with a signature function $\sigma: \cE \rightarrow \{-1,+1\}$ that associates a sign to each edge.  Each edge $(u,v)\in \cE$ can also be assigned a nonnegative weight $w_{uv} \geq 0$.  The degree of a vertex is
$$d_u :=\sum_{v: (u,v) \in \cE} w_{uv},$$
which does not depend on $\sigma$.  For a subset $S\subseteq(\cV)$ define $\vol(S) = \sum_{v \in \
\cV} d_v$.  Define $L^\sigma := D - A^\sigma$, where
$$
D := \sum_{u \in \cV} d_u |u\rangle \langle u| \quad , \quad A^\sigma := \sum_{(u,v) \in \cE} \sigma_{uv} w_{uv}|u\rangle \langle v| ,
$$
\noindent which are the diagonal degree matrix, the signed adjacency matrix, and the signed graph Laplacian.  Denote the eigenvectors of $L^\sigma$ by $\{\phi^\sigma_i\}_{i=1}^n$ with the corresponding eigenvalues,
$$
0 \leq \lambda_1^\sigma\leq \lambda_2^\sigma \leq ... \leq \lambda_n^\sigma  \leq 2 D_{\max} .
$$
In the special case of the all-positive signature, i.e., $\sigma_{uv} = +1$ for all $(u,v) \in E$, the corresponding Laplacian $L^+$ is the standard graph Laplacian of $G$ and is an explicitly stoquastic Hamiltonian.  In general $L^\sigma$ can be non-stoquastic, and in these cases its stoquastization corresponds to $L^+$.  We've already seen the bound
$$
\lambda_1^+ \leq \lambda^\sigma_1
$$
(and in fact $\lambda^+_1 = 0$ for all unsigned graph Laplacians).  Therefore it remains to find relations between the non-stoquastic first excited energy $\lambda^\sigma_2$ in terms of the spectrum of $L^+$, and our original results of this nature are derived in Section~\ref{sec:newresults}.  
\subsection{Background} \label{sec:background}
First we describe the important class of balanced graphs, for which the sign problem in the signed graph Laplacian is curable.  Then we define the frustration index, which is a quantitative measure of how far a graph is from being balanced.   Then we review Cheeger inequalities for unsigned graphs, in order to appreciate how they become modified by the frustration index to obtain the signed Cheeger inequalities.

\paragraph{Balanced graphs.} Let $\theta$ be a diagonal matrix with $\pm 1$ along the diagonal.  Taking $L^\sigma$ to $\theta^{-1} L^\sigma \theta$ is a unitary transformation, and so in quantum mechanics we view it as a change of basis.  Graph theorists called $\theta$ a switching function, and the new signed graph defined by $\theta^{-1} A^\sigma \theta = A^{\sigma'}$ is called switching-equivalent to the original.  Needless to say, switching preserves the spectrum.  Equivalence under switching defines a partition of the set of signatures into equivalence classes $[\sigma]$.  A signed graph in the same switching class as the all-positive signature is called \emph{balanced}.  It has the property that the product of the edge signatures around every cycle is positive.  This is the equivalence class of Hamiltonians that have a sign problem that is curable in the vertex label basis.

It turns out we will need another characterization of balance throughout the following sections.  A graph is balanced if and only if there exists a bipartition $\cV = V_1 \cup V_2$, with $V_1 \cap V_2 = \emptyset$, such that each positive edge connects to elements of the same subset, and each negative edge connects elements of opposite subsets.  We call this a balanced bipartition. For the graph with all-positive signature this is achieved by taking $V_1 = \cV$ and $V_2 = \emptyset$.  In general the bipartition may be nontrivial.

\paragraph{Frustration Index.} 
In the previous section we saw that a graph is balanced iff it admits a balanced bipartition.  In this section we define the frustration index of an arbitrary subset of vertices, which captures the residual frustration in the most-balanced possible bipartition of the subset.  To define the most-balanced bipartition we need some additional notation.

Let $V_1,V_2 \subseteq \cV$ and define functions that count the number of edges between $V_1$ and $V_2$ of positive, negative, or either type:
\begin{align}
|\cE^+(V_1,V_2)| &:= \sum_{\substack{(u,v) \in \cE, \; \sigma_{uv} = +1 \\ u \in V_1, v \in V_2 }} w_{uv} , \\ |\cE^-(V_1,V_2)| &:= \sum_{\substack{(u,v) \in \cE, \; \sigma_{uv} = -1 \\ u \in V_1, v \in V_2 }} w_{uv},
\end{align}
and $|\cE(V_1,V_2)| = |\cE^+(V_1,V_2)| + |\cE^-(V_1,V_2)|$.  define $|\cE^+(V_1)| = |\cE^+(V_1,V_1)|$ and similarly for $|\cE^-(V_1)|$ and $|\cE(V_1)|$.  Note that each edge is counted twice in these sums.
Now for any subset $S \subseteq \cV$ define the frustration index $F(S)$, which captures the minimum residual energy due to non-stoquastic frustration in that subset,
\begin{align}
F(S) := \frac{1}{\vol(S)} \cdot \min_{\substack{S_1,S_2 \subseteq S \\S_1 \cap S_2 = \emptyset }} \bigl (2 |\cE^+(S_1,S_2)|\nonumber \\ + |\cE^-(S_1)| + |\cE^-(S_2)| \bigr ).\label{eq:frustrationIndex}
\end{align}
It turns out that $F(S)$ is also related to the minimum number of edges which need to be removed from the subgraph induced on $S$ in order to put it in the balanced signature equivalence class.

\paragraph{(Unsigned) Cheeger inequalities.}  
\begin{figure}
\includegraphics[width=0.4 \textwidth]{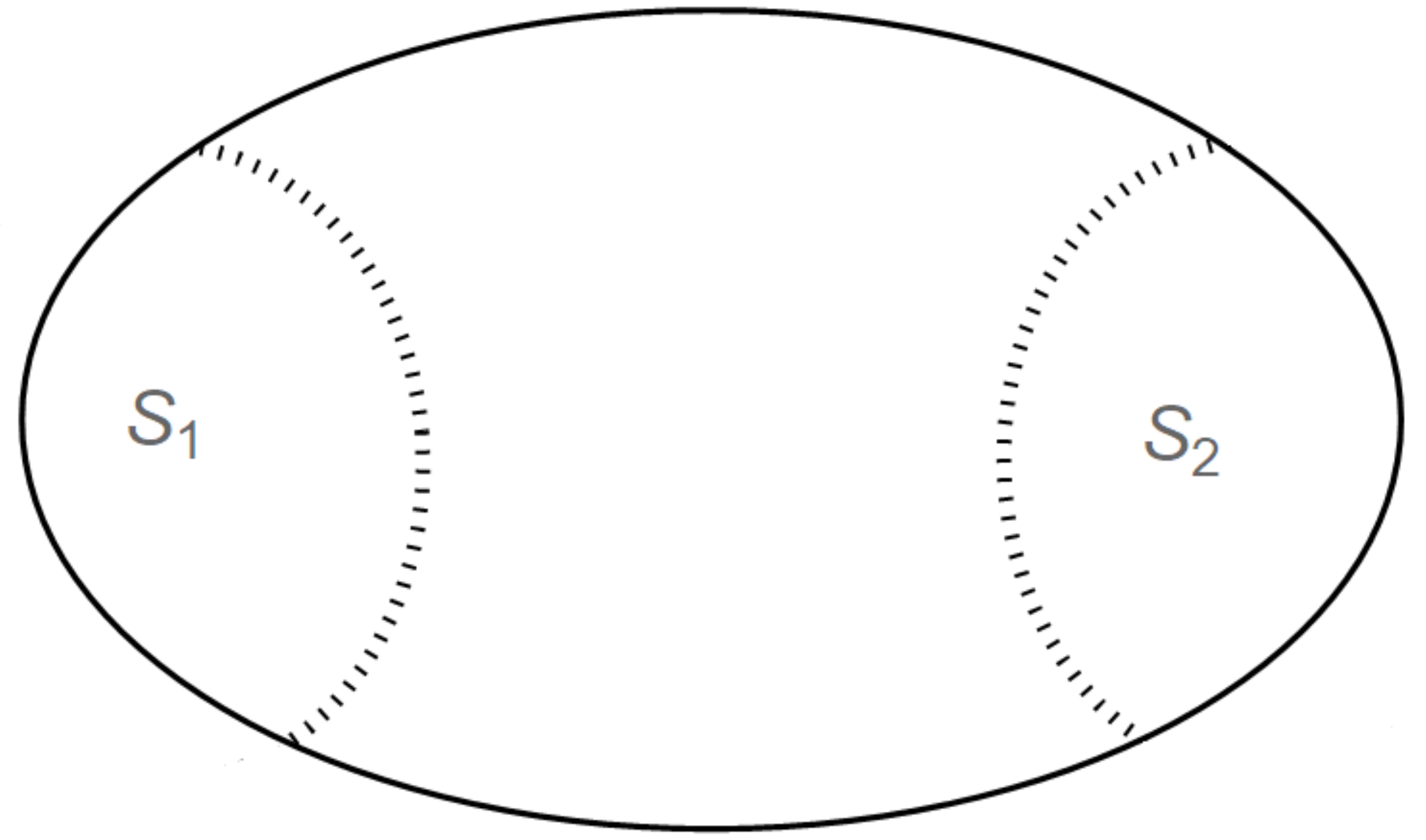}
\includegraphics[width=0.4 \textwidth]{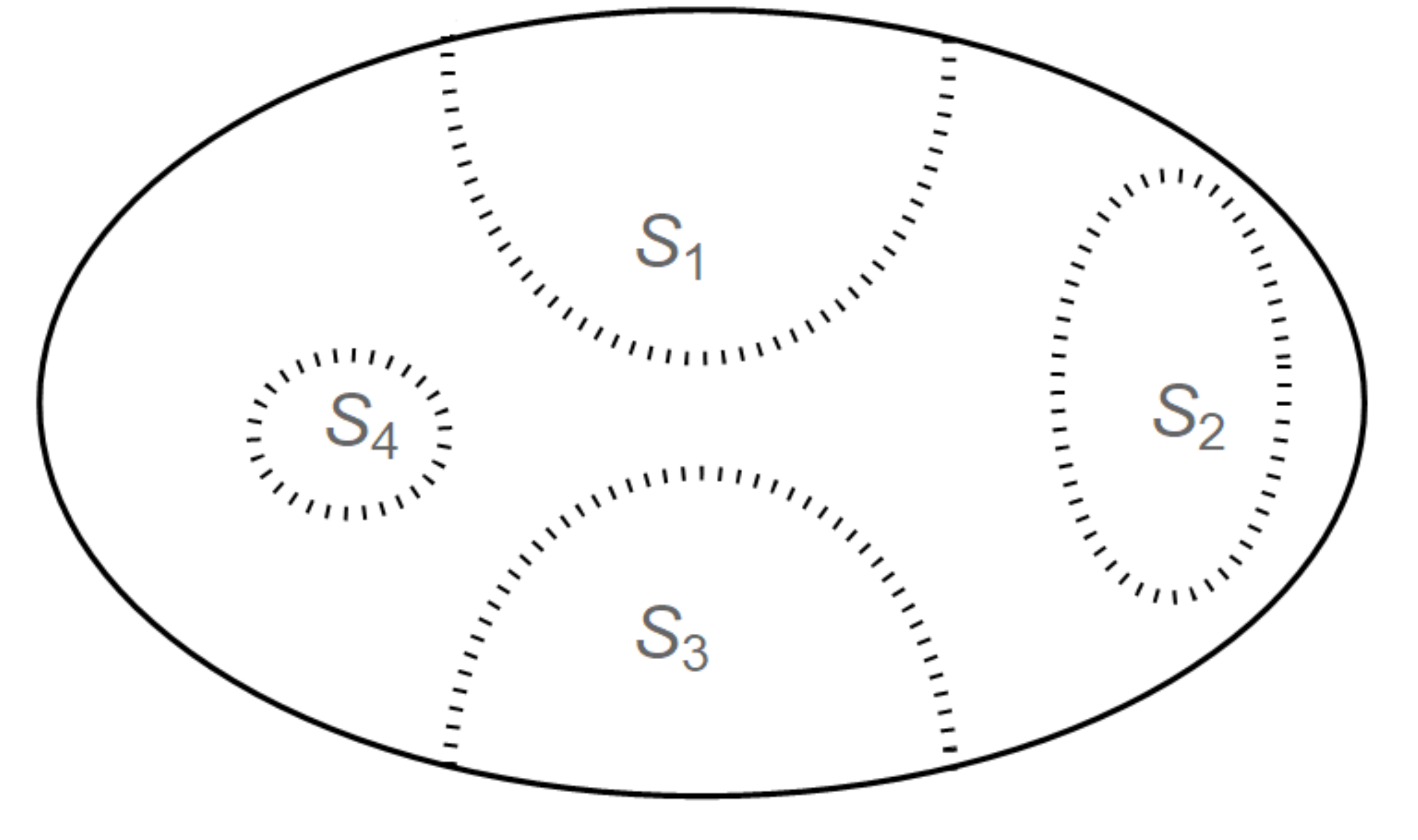}
\caption{\label{fig:UnsignedCheeger}Top: An illustration of a graph as a geometric space (vertices and edges are not explicitly shown) with two disjoint subsets $S_1$, $S_2$ which could achieve the minimum in Eq.~\eqref{eq:stoqCheegerDef}. Bottom: An example of 4 subsets that could determine $h_4^+$ as in Eq.~\eqref{eq:multiStoqCheegerDef}.  Using the higher-order Cheeger inequality in Eq.~\eqref{eq:multiStoqCheeger} these 4 sets could determine the 3 lowest eigenvalues of $L^+$.  }
\end{figure}

For any subset $S \subseteq \cV$ the \emph{expansion} is defined by
$$
\Phi(S) = \frac{|\cE(S,\bar{S})|}{\vol(S)} \ .
$$ This quantity is also sometimes called the bottleneck ratio, because a small value of $\Phi(S)$ indicates a relatively large number of vertices and small number of edges, corresponding to a bottleneck in the graph.

For an unsigned graph the following Cheeger constant captures the minimum possible expansion over all disjoint pairs of subsets, 
\begin{equation}
h^+_2= \min_{\substack{S_1,S_2= \cV\\S_1 \cap S_2 = \emptyset}}  \max \left \{\Phi(S_1), \Phi(S_2) \right\}. \label{eq:stoqCheegerDef}
\end{equation}
Taking the maximum expansion of two disjoint subsets ensures that at least one of them has $\vol(S_i) \leq \vol(\cV)/2$.   

Cheeger's inequality relates Eq.~\eqref{eq:stoqCheegerDef} to $\lambda_2^+$ (which is the spectral gap because $\lambda^+_1 = 0$),
\begin{equation}
\frac{\lambda^+_2}{2} \leq h^+_2 \leq \sqrt{2 D_{\max} \lambda_2^+}. \label{eq:unsignedCheeger}
\end{equation}

The proof for the bound $\lambda^+_2 \leq 2 h^+_2$ is that for any $S$ we can take a first excited state ansatz that is positive on $S$ and negative on $\bar{S}:= \cV - S$,
$$
\hat{\phi}_2^+(x) = \begin{cases} 
      \vol(\bar{S}) & , x\in S \\
      -\vol(S) &, x \in \bar{S} 
   \end{cases} .
$$ 
The upper bound $h^+_2 \leq \sqrt{2 D_{\max} \lambda_2^+}$ says that if the gap is small, then there is a subset with small expansion.  The proof is based on a spectral clustering algorithm.  The idea is that the first excited state can be divided into vertices with positive amplitude and vertices with negative amplitude.  Then these sets are further refined a bit to obtain the subsets $S_1, S_2$.  

The Cheeger constant and it's relation to the first excited energy can also be generalized to the rest of the spectrum by what are called higher-order Cheeger inequalities. Define the $k$-th order Cheeger constant to be
\begin{equation}
h^+_k= \min_{\substack{\bigcup_{i=1}^k S_i =  \cV\\S_i \cap S_j = \emptyset}}  \max \left \{\Phi(S_1), ... , \Phi(S_k) \right\}. \label{eq:multiStoqCheegerDef}
\end{equation}
which are related to the $k$-eigenvalue,
\begin{equation}
\frac{\lambda^+_k}{2} \leq h_k^+ \leq C k^3 \sqrt{2 D_{\max} \lambda_k^+},\label{eq:multiStoqCheeger}
\end{equation}
where $C$ is a constant.   Figure~\ref{fig:UnsignedCheeger} illustrates a sub-partition with 4 subsets which can be used to upper bound $h_4^+$ and hence upper bound $\lambda_k^+$.  
\paragraph{Signed Cheeger inequalities.}

\begin{figure}
\includegraphics[width=0.4 \textwidth]{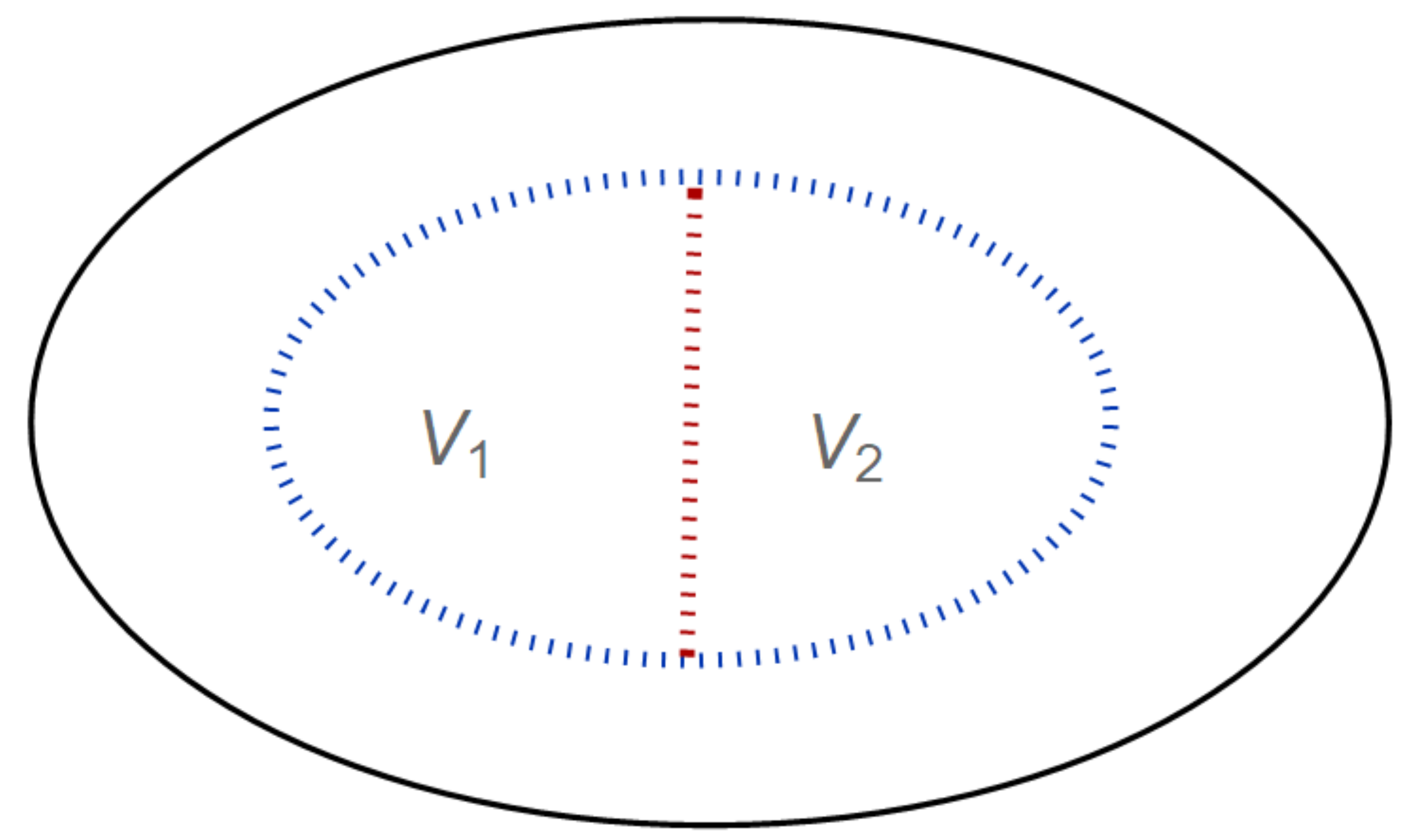}
\includegraphics[width=0.4 \textwidth]{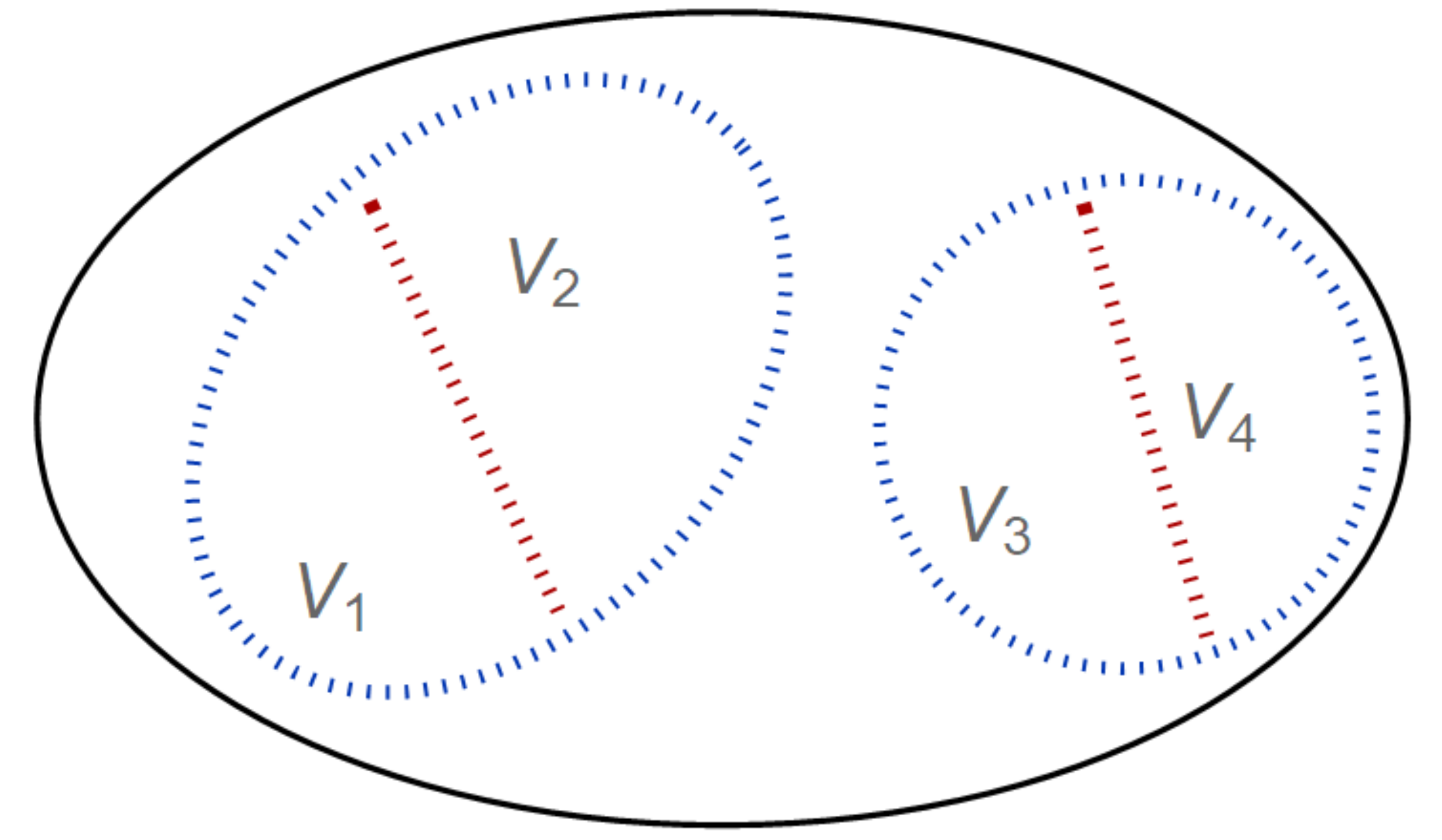}
\caption{Top: The set $S = V_1 \cup V_2$ that simultaneously minimizes its frustration index $F(S)$ and its edge expansion $\Phi(S)$ determines the signed Cheeger constant in Eq.~\eqref{eq:signedCheegerDef} and the ground energy in Eq.~\eqref{eq:signedCheeger}. Bottom: Two subsets $S_1 = V_1 \cup V_2$ and $S_2 = V_3 \cup V_4$ that simultaneously minimize frustration and edge expansion.  These sets could determine the first excited energy $\lambda_2^\sigma$ according to Eq.~\eqref{eq:higherSignedCheeger}.  }
\label{fig:UnsignedHigherCheeger2}
\end{figure}
The lowest order signed Cheeger constant is defined to be
\begin{equation}
h_1^\sigma = \min_{S \subseteq \cV} F(S) + \Phi(S). \label{eq:signedCheegerDef}
\end{equation}
The meaning of this quantity is that it seeks to find the subset of vertices in the graph which is simultaneously close to balanced and has small expansion.  Such a subset can be used to construct a good ansatz for the ground state, which is nonzero in $S$ and vanishes outside of $S$.  The most-balanced bipartition $S = V_1 \cup V_2$ is then used to assign the ansatz positive values in $V_1$ and negative values in $V_2$.  This leads to the lowest order signed Cheeger inequality,
\begin{equation}
\frac{\lambda_1^\sigma}{2} \leq h_1^\sigma \leq \sqrt{2 D_{\max} \lambda_1^\sigma} \label{eq:signedCheeger}
\end{equation}
It is also possible to define a series of higher order signed Cheeger constants which bound the eigenvalues above the ground state of $L^\sigma$, 
\begin{equation}
h^\sigma_k= \min_{\substack{\bigcup_{i=1}^k S_i = \cV\\S_i \cap S_j = \emptyset}}  \max \left \{F(S_1)  +\Phi(S_1) , ... , F(S_k) + \Phi(S_k) \right\}. \label{eq:higherCheegerDef}
\end{equation}
which for any signature $\sigma$ satisfy
\begin{equation}
\frac{\lambda^\sigma_k}{2} \leq h_k^\sigma \leq C k^3 \sqrt{2 D_{\max} \lambda_k^\sigma}.\label{eq:higherSignedCheeger}
\end{equation}
In particular we will use the case $k = 2$ (for which the definition~\ref{eq:higherCheegerDef} is illustrated in Fig.~\ref{fig:UnsignedHigherCheeger2}) to upper $\lambda^\sigma_2$ and hence upper bound the non-stoquastic spectral gap.
\subsection{New Results}\label{sec:newresults}

\begin{figure}
\includegraphics[width=0.4\textwidth]{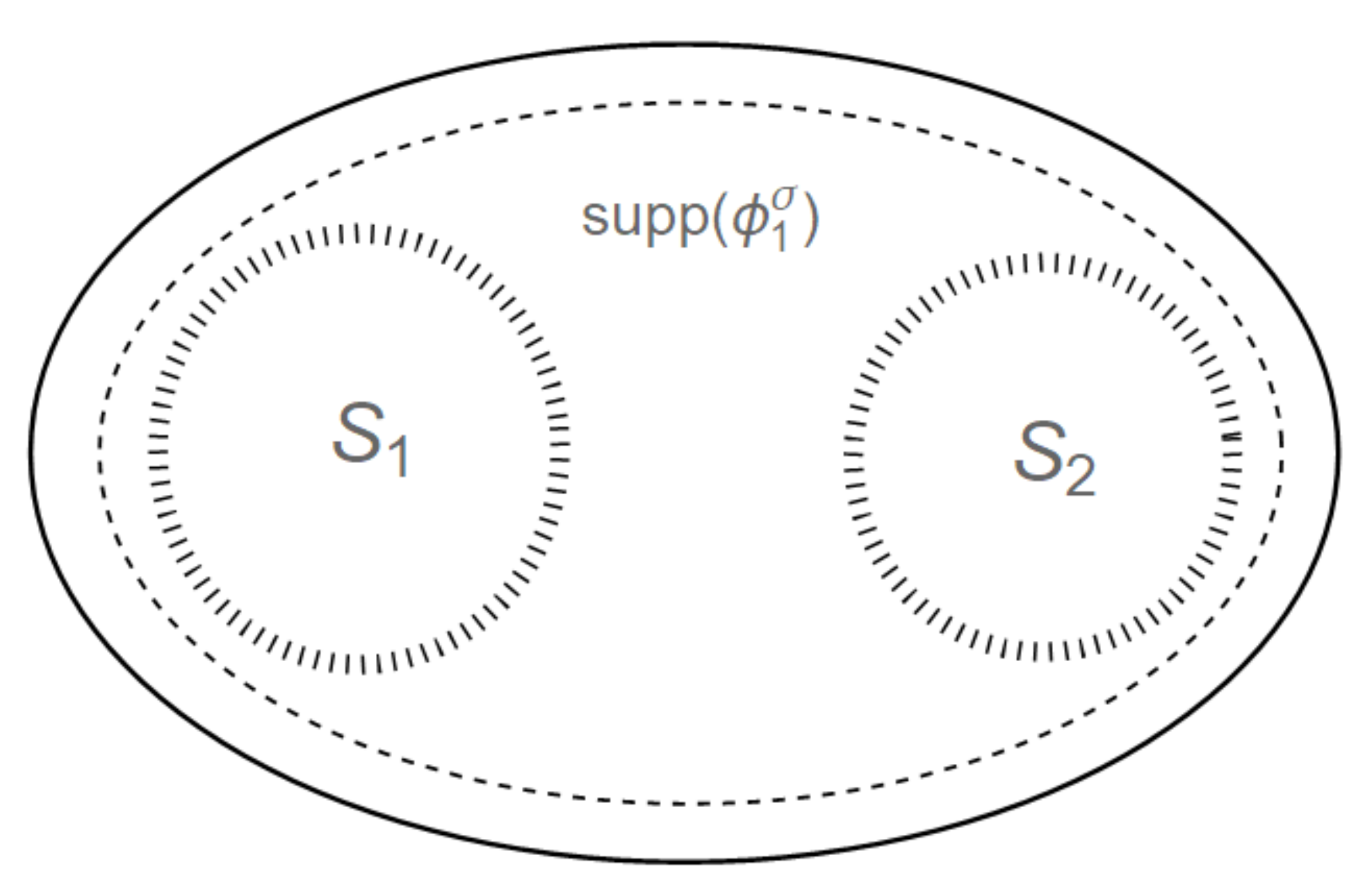}
\includegraphics[width=0.4\textwidth]{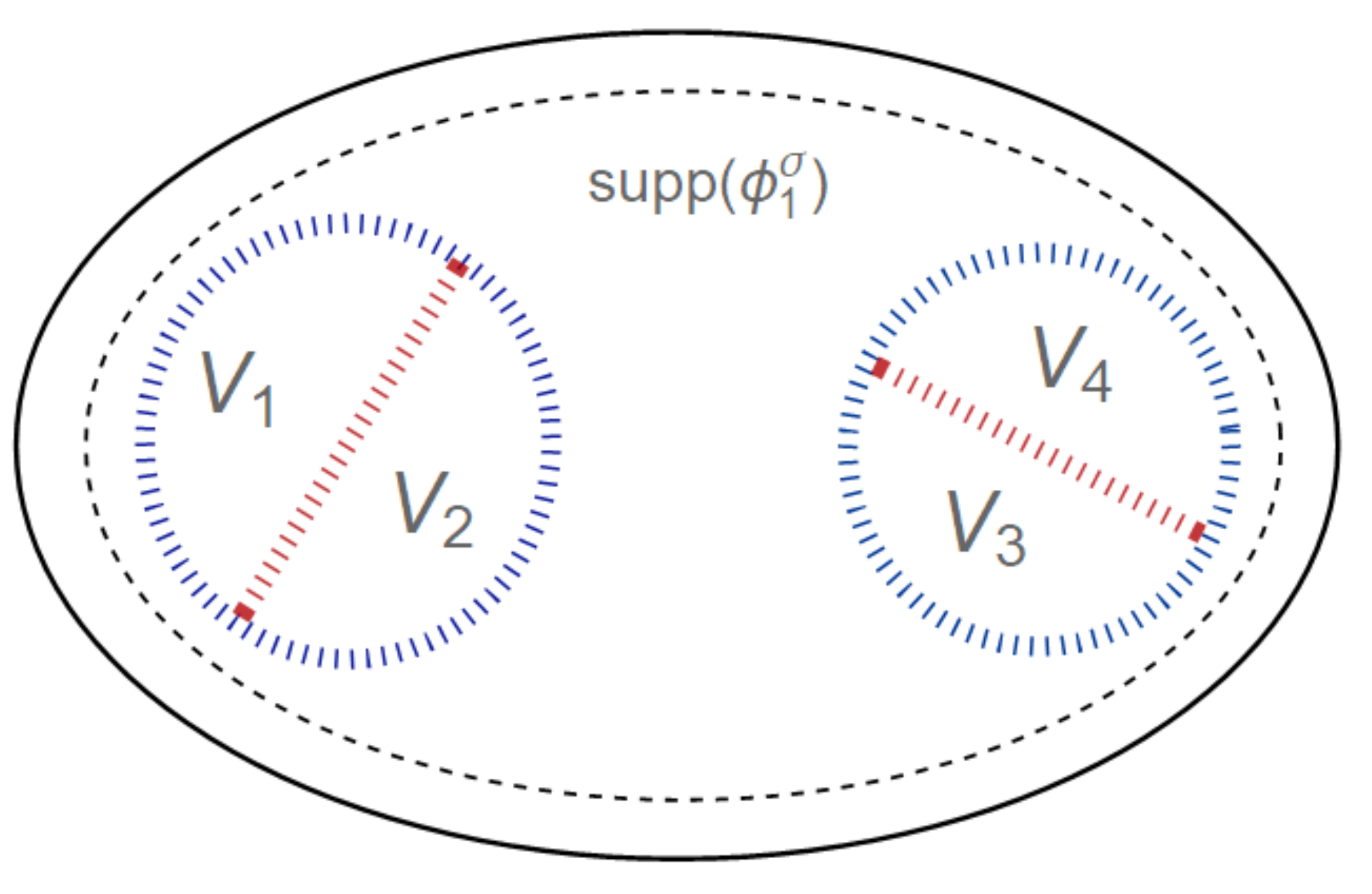}
\caption{Top: The interior region $\Omega$ is the support of the ground state of $L^\sigma$.  The sets $S_1,S_2$ are Cheeger subsets for the unsigned Laplacian $L^+_{\Omega}$ restricted to the subgraph $\Omega$.  Using Eq.~\eqref{eq:unsignedCheeger} the edge expansions $\Phi(S_1),\Phi(S_2)$ are upper bounded in terms of $\lambda^+_{2,\Omega}$. Bottom: The most-balanced bipartitions $S_1 = V_1 \cup V_2$ and $S_2 = V_3 \cup V_4$ are used to upper bound $\lambda^\sigma_2$ with Eq.~\eqref{eq:higherSignedCheeger}.  The frustration index of these sets is upper bounded by $\frac{\vol(\Omega)}{\vol(S_i)}\sqrt{\lambda_1^\sigma}$ by Eq.~\eqref{eq:signedCheeger} , and the edge expansion is upper bounded by $\sqrt{\lambda_{2,\Omega}^+}$ by Eq.~\eqref{eq:unsignedCheeger}.}\label{fig:UnsignedHigherCheeger3}
\end{figure}

Let $\Omega \subseteq \cV$ be the support of the ground state of the signed Laplacian,
$$
\Omega := \left \{v \in \cV : \phi^\sigma_1(v) \neq 0 \right \},
$$
The first bound we present is given in terms of the first excited energy of the unsigned graph $\Omega$, considered as a subgraph of $G$.  The proof strategy is illustrated in Fig.~\ref{fig:UnsignedHigherCheeger3}. This unsigned subgraph Laplacian is $L^+_{\Omega}$, and the eigenvalue and eigenvector are $\lambda^+_{2,\Omega}$, $\phi_{2,\Omega}^+$.  By Cheeger's inequality we know there exists sets $S_1,S_2$ with 
$$
\max \left \{\Phi(S_1), \Phi(S_2) \right\} \leq \sqrt{2 D_{\max} \lambda^+_{2,\Omega}}
$$
Now we use $S_1,S_2$ to upper bound the second eigenvalue of unsigned Laplacian
\begin{align*}
\lambda^\sigma_2 &\leq 2 \max \left \{F(S_1) + \Phi(S_1), F(S_2) + \Phi(S_2) \right\} \\
&\leq 2 \max \left \{F(S_1),F(S_2) \right \} + \sqrt{2 D_{\max}  \lambda^+_{2,\Omega}}.
\end{align*}
In order to upper bound $F(S_1),F(S_2)$ we need the fact that for any subsets $S\subseteq S'$ we have 
$$
F(S) \leq \frac{\vol (S')}{\vol (S)} F(S').
$$
This is a comparison of the numerators in Eq.~\eqref{eq:frustrationIndex}, which states that the most-balanced bipartition does not get better as more vertices and signed edges are added.   Therefore
\begin{align*}
\lambda^\sigma_2 &\leq 2 \vol (\Omega) F(\Omega) \max \left \{\frac{1}{\vol (S_1)}, \frac{1}{\vol (S_2)} \right\} \\
&+ \sqrt{2 D_{\max} \lambda^+_{2,\Omega}}.
\end{align*}
and this leads to the following theorem.
\begin{theorem}
\textit{ Let $L^\sigma = D - A^\sigma$ be a signed graph Laplacian with ground state $\phi_1^\sigma$ with energy $\lambda^\sigma_1$.  Define $\Omega := \{v \in V : \phi_1^\sigma(v) \neq 0 \}$ and consider the unsigned Laplacian  $L^+_\Omega$ on the subgraph $\Omega$.  By Cheeger's inequality there must be a subset $S \subseteq \Omega$ with $\Phi(S) \leq \sqrt{2 D_{\max} \lambda^+_{2,\Omega}}$, and for any subset $S$ with expansion $\Phi(S)$ satisfying this relation we have }
$$
\lambda^\sigma_2 \leq 2\sqrt{2 D_{\max}}\left( \frac{\vol(\Omega)}{\vol(S)}\sqrt{\lambda^{\sigma}_1} + \sqrt{\lambda^+_{2,\Omega}}  \right),
$$
\textit{where $\lambda^+_{2,\Omega}$ is the first excited energy of the unsigned subgraph Laplacian $L^+_\Omega$.  }
\end{theorem}

Using the same proof idea we can also obtain an upper bound directly in terms of $\lambda^+_2$ without restricting to the subgraph $\Omega$, at the cost of introducing more volume factors.  This time the subsets $S_1, S_2$ come from applying Cheeger's inequality to the full unsigned Laplacian $L^+$, and then forming the intersections $R_1 = S_1 \cap \Omega$ and $R_2 = S_2 \cap \Omega$ to take the variational upper bound 
$$
\lambda^\sigma_2 \leq  2 \max \left \{F(R_1) + \Phi(R_1), F(R_2) + \Phi(R_2) \right\}.
$$ 
As before we can upper bound the frustration
$$
F(S) \leq \frac{\vol(\Omega)}{\vol(R)} F(\Omega) \leq \frac{\vol(\Omega)}{\vol(R)}\sqrt{2 D_{\max} \lambda^\sigma_1}
$$
where $R$ is either $R_1$ or $R_2$.  Next observe that every vertex on the boundary of $R = S \cap \Omega$ is either on the boundary of $S$ or the boundary of $\Omega$, and so
\begin{align*}
\Phi(R) &\leq  \frac{\vol(\Omega)}{\vol(R)} \Phi(\Omega) + \frac{\vol(S)}{\vol(R)} \Phi(S) 
 \\
& \leq \frac{\vol(\Omega)}{\vol(R)} \sqrt{2 D_{\max}  \lambda^\sigma_1} +
\frac{\vol(S)}{\vol(R)} \sqrt{2 D_{\max}  \lambda^+_2} 
\end{align*}
Therefore we have now shown the following variant of the theorem.
\begin{theorem}
\textit{ Let $L^\sigma = D - A^\sigma$ be a signed graph Laplacian with ground state $\phi_1^\sigma$ with energy $\lambda^\sigma_1$.  Define $\Omega := \{v \in V : \phi_1^\sigma(v) \neq 0 \}$ and consider the unsigned Laplacian $L^+$ on the same graph.  By Cheeger's inequality there must be a subset $S \subseteq V$ with $\Phi(S) \leq \sqrt{2 D_{\max}  \lambda^+_{2}}$, and for any subset $S$ with expansion $\Phi(S)$ satisfying this relation we have }
$$
\lambda^\sigma_2 \leq 2\sqrt{2D_{\max} }\left( \frac{\vol(\Omega)}{\vol(S \cap \Omega)}\sqrt{\lambda^{\sigma}_1} + \frac{2 \; \vol(S)}{\vol(S \cap \Omega)}\sqrt{\lambda^+_{2}}  \right)
$$
\end{theorem}

Finally we present a cleaner theorem that unfortunately produces a statement that goes in the opposite direction from what we want.

\paragraph{Theorem 3 (converse bounds):} For any graph $G$ and any signature $\sigma$, we have
$$
\lambda_2^+ \leq \sqrt{2 D_{\max} \lambda^\sigma_2}
$$
\paragraph{Proof.}
\begin{align*}
\lambda^+_2 &\leq 2h^+_2\\
&= \min_{\substack{S_1, S_2\subseteq V\\S_1 \cap S_2 = \emptyset}}  \max \left \{\Phi(S_1), \Phi(S_2) \right\}\\
&\leq \min_{\substack{S_1, S_2\subseteq V\\S_1 \cap S_2 = \emptyset}}  \max \left \{F(S_1) + \Phi(S_1), F(S_2) + \Phi(S_2) \right\}\\
& = h^\sigma_2\\
&\leq \sqrt{2 D_{\max}  \lambda^\sigma_2}
\end{align*}
\paragraph{Comments.}  This shows that the first excited energy of the unsigned graph cannot be too much larger than the first excited energy of the signed graph.  However since $\lambda^+_1 \leq \lambda^\sigma_1$, the unsigned graph could still have a larger spectral gap.

\subsection{How tight are these bounds?}
In the case of theorem 1, the only volume factor is $\vol(\Omega)/\vol(S)$.  Here $S$ is a Cheeger subset for a graph defined on $\Omega$.  Usually the Cheeger subset occupies about half of the vertices, for example in a path graph $\{1,...,k\}$ the subset $S = \{1, ... , \lfloor k/2 \rfloor\}$ achieves the minimum expansion.  Or in an $n$-dimensional hypercube, we would take a Hamming ball of radius $\lfloor n/2 \rfloor.$  So in many cases we expect $\vol(\Omega)/\vol(S) = \mathcal{O}(1)$ and we'll have
$$
\lambda^\sigma_2 = \mathcal{O} \left( \sqrt{D_{\max} \lambda^\sigma_1} +\sqrt{D_{\max} \lambda^+_{2,\Omega}} \right).
$$
For theorem 2 in the previous section, the RHS can blow up when $S \cap \Omega = \emptyset$, and this does in fact sometimes happen.  What this means is that the first excited state in stoquastic case is very different from the non-stoquastic ground state.  In other words, the subset that minimizes the expansion does a very poor job at minimizing the phase frustration.  The best case for theorem 2 occurs when $\Omega = V$, and theorem 1 and 2 coincide.  The only advantage theorem 2 ever has is that it refers to the eigenvalue $\lambda^+_2$ instead of $\lambda^+_{2,
\Omega}$.

\paragraph{Improved Cheeger inequalities and the square root.}

The topic of improved Cheeger inequalities is based on using higher eigenvalues to obtain tighter bounds.   For any signed graph and any $k \in \{1,...,N\}$ we have~\cite{atay2014cheeger}
$$
h_1^\sigma \leq 16 \sqrt{2 \| D\|} k \frac{\lambda_1^\sigma}{\sqrt{\lambda^\sigma_k}}.
$$
The key point being the absence of the square root on $\lambda^\sigma_1$ on the RHS.  The meaning of this inequality is that we can get a quadratically improved connection between expansion and gap, in the cases when there one of the higher eigenvalues (for $k \geq 2$ but not too large because of the $k$ on the RHS above) is much larger than the bare spectral gap.  

For example, consider a transverse Ising model in the ferromagnetic regime.  It has two exponentially close eigenvalues, then a constant gap to the rest of the spectrum.  The improved Cheeger inequality tells us that the expansion and the gap scale with the same order in this ground state.

An interesting thing about this is that it is similar to the condition under which diabatic QA succeeds: the low energy spectrum (which may have polynomially many states, say) is separated from the rest of the spectrum by a gap.  There is also higher order improved signed Cheeger inequality, which says that there exists a constant C such that for any signed graph and any $1 \leq k \leq l \leq N$,
$$
h_k^\sigma \leq C \sqrt{\|D\|} l k^6 \frac{\lambda_k^\sigma}{\sqrt{\lambda^\sigma_l}}
$$
Therefore, if QA succeeds then there should be a gap after $k = \poly(n)$ eigenvalues, and so we expect that in many cases the bound tightens to
$$
\lambda^\sigma_2 = \mathcal{O} \left(\poly(n) \left( \lambda^\sigma_1 +\lambda^+_{2,\Omega} \right )\right).
$$

\end{document}